\documentclass[12pt]{article}%
\usepackage{amsfonts}
\usepackage{amssymb,mathrsfs}
\usepackage[fleqn]{amsmath}
\usepackage{lmodern}
\usepackage{geometry}
\usepackage{graphicx}
\usepackage{setspace}
\usepackage{booktabs}
\providecommand{\U}[1]{\protect\rule{.1in}{.1in}}
\newtheorem{theorem}{Theorem}

\newtheorem{lemma}[theorem]{Lemma}

\newtheorem{proposition}[theorem]{Proposition}
\newtheorem{remark}[theorem]{Remark}

\newcommand{\mP}{\mathbb{P}}
\newcommand{\Pihat}{\widehat{\Pi}}

\usepackage{amsmath,amssymb}
\usepackage{epsfig,ifthen,
	amsmath,amssymb,ifthen,
	pstricks,pst-node}
\usepackage{soul,color}
\definecolor{mygreen}{RGB}{144,241,47}

\usepackage{booktabs}
\usepackage{epstopdf}
\DeclareGraphicsExtensions{.ps}
\newcounter{isamac} %counter used to indicate if compiled on mac
\ifthenelse{\value{isamac}=1}{
\input BoxedEPS          %
\SetTexturesEPSFSpecial  % Optional Mac graphics commands
\HideDisplacementBoxes   %
}{}

\newcommand{\indm}[2]{\ensuremath{{\mathfrak I}_{\kern-1pt\scriptstyle#1}({\mathcal
#2})}} % for independence models

\newcommand{\ind}{\mbox{$\perp \kern-5.5pt \perp$}}

\newcommand{\uned}{\hbox{\kern3pt\raise2.5pt\vbox{\hrule
width9pt height 0.3pt}\kern3pt}}

\newcommand{\dashed}{\hbox{\kern3.05pt\raise2.5pt\vbox{\hrule%dashed edges
width1.7pt height 0.3pt}\kern1.8pt\raise2.5pt\vbox{\hrule
width1.7pt height 0.3pt}\kern1.8pt\raise2.5pt\vbox{\hrule
width1.7pt height 0.3pt}\kern1.8pt\raise2.5pt\vbox{\hrule
width1.7pt height 0.3pt}\kern3.05pt}}

\newcommand{\pedg}[2]{\ensuremath{{\kern0.5pt
\scriptstyle{\ifthenelse{\equal{\head}{#1}}{\lhead\kern0.5pt}{#1\kern0.5pt}}\joinrel\relbar
\negthinspace\relbar\joinrel{\kern0.5pt #2}\kern0.5pt}}}
\newcommand{\pdots}{\hbox{\kern2.5pt\raise1.5pt\hbox{\ensuremath{\ldots}}\ke
rn2.5pt}}  %between PAG edges

\usepackage{algorithm}
\usepackage{setspace}
\usepackage{times}

\usepackage{tikz}
\usetikzlibrary{arrows,shapes.arrows,shapes.geometric,
	shapes.multipart,backgrounds,decorations.pathmorphing,positioning,fit,automata}
\tikzset{
	>=stealth',
	true/.style={
		rectangle,
		draw=black, very thick,
		text width=6.5em,
		minimum height=2em,
		text centered,
		fill=gray, opacity = 0.5},
	punkt/.style={
		rectangle,
		rounded corners,
		draw=black, very thick,
		text width=6.5em,
		minimum height=2em,
		text centered},
	est/.style={
		circle,
		draw=black, very thick,
		text width=4em,
		minimum height=2em,
		text centered},
	weight/.style={
		circle,
		draw=black, very thick,
		text width=6.5em,
		minimum height=2em,
		text centered},
	pil/.style={
		->,
		thick,
		shorten <=2pt,
		shorten >=2pt,},
	double/.style={
		<->,
		thick,
		shorten <=2pt,
		shorten >=2pt,},
	dash/.style={
		dashed,
		thick,
		shorten <=2pt,
		shorten >=2pt,},
	dashdouble/.style={
		<->,
		dashed,
		thick,
		shorten <=2pt,
		shorten >=2pt,}
}

\usepackage{array}
\usepackage[titletoc,title]{appendix}
\usepackage{amssymb,amsfonts,latexsym,graphicx}
\usepackage{color}

\usepackage{wasysym,MnSymbol}%
\usepackage{bm}
\usepackage{natbib}
\renewcommand{\ind}{\Large\mbox{$\perp \kern-5.5pt \perp$}}

\newcommand\numberthis{\addtocounter{equation}{1}\tag{\theequation}}
\newcommand{\fss}{full subclassification }
\newcommand{\fsw}{(full) subclassification }
\newcommand*{\ps}{propensity score }
\newcommand{\HT}{Horvitz-Thompson }

\newtheorem{assumption}{Assumption}
\usepackage{color}
\definecolor{orange}{rgb}{1,0.5,0}
\definecolor{gray}{gray}{0.5}

\newcommand{\hajek}{\hat{\Delta}_{Ratio}}

\begin{document}

%%%%%%%%%%%%%%%%%%%%%%%%%%%%%%%%%%%%%%%%%%%%%%%%%%%%%%%%%%%%%%%%%%%%%%%%%%%%%%%%%%%%%%%%%%%%%%%%%%%%%%%%%%%%%%%%%%%%%%%%%%%%
%%%%%%%%%%%%%%%%%%%%%%%%%%%%%%%%%%%%%%%%%%%%%%%%%%%%%%%%%%%%%%%%%%%%%%%%%%%%%%%%%%%%%%%%%%%%%%%%%%%%%%%%%%%%%%%%%%%%%%%%%%%%

\renewcommand{\baselinestretch}{2}

\markright{ \hbox{\footnotesize\rm 
%{\footnotesize\bf 24} (201?), 000-000
}\hfill\\[-13pt]
\hbox{\footnotesize\rm
%\href{http://dx.doi.org/10.5705/ss.20??.???}{doi:http://dx.doi.org/10.5705/ss.20??.???}
}\hfill }

\markboth{\hfill{\footnotesize\rm LINBO WANG AND YUEXIA ZHANG AND THOMAS S. RICHARDSON AND XIAO-HUA ZHOU} \hfill}
{\hfill {\footnotesize\rm ROBUST ESTIMATION OF PROPENSITY SCORE WEIGHTS} \hfill}

\renewcommand{\thefootnote}{}
$\ $\par

%%%%%%%%%%%%%%%%%%%%%%%%%%%%%%%%%%%%%%%%%%%%%%%%%%%%%%%%%%%%%%%%%%%%%%%%%%%%%%%%%%%%%%%%%%%%%%%%%%%%%%%%%%%%%%%%%%%%%%%%%%%%

\fontsize{12}{14pt plus.8pt minus .6pt}\selectfont \vspace{0.8pc}
\centerline{\large\bf ROBUST ESTIMATION OF PROPENSITY SCORE WEIGHTS }
\vspace{2pt} \centerline{\large\bf VIA SUBCLASSIFICATION }
\vspace{.4cm} \centerline{$^{1}$Linbo Wang, $^{2}$Yuexia Zhang, $^{3}$Thomas S. Richardson, $^{4}$Xiao-Hua Zhou} \vspace{.4cm} \centerline{\it $^{1,2}$University of Toronto, $^{3}$University of Washington, $^{4}$Peking University
} \vspace{.55cm} \fontsize{9}{11.5pt plus.8pt minus
.6pt}\selectfont

%%%%%%%%%%%%%%%%%%%%%%%%%%%%%%%%%%%%%%%%%%%%%%%%%%%%%%%%%%%%%%%%%%%%%%%%%%%%%%%%%%%%%%%%%%%%%%%%%%%%%%%%%%%%%%%%%%%%%%%%%%%%

\begin{quotation}
\noindent {\it Abstract:}
Weighting estimators based on propensity scores are widely used for causal estimation in a variety of contexts, such as observational studies, marginal structural models and  interference. They enjoy appealing theoretical properties such as consistency and possible efficiency under correct model specification. However, this theoretical appeal may be diminished in practice by sensitivity to misspecification of the propensity score model. To improve on this, we borrow an idea from an alternative approach to causal effect estimation in observational studies, namely subclassification estimators. It is well known that compared to weighting estimators,  subclassification methods are usually more robust to model misspecification. In this paper, we first discuss an intrinsic  connection between the seemingly unrelated weighting and subclassification estimators, and then use this connection to construct robust propensity score weights via subclassification. We illustrate this idea by proposing so-called full-classification weights and accompanying estimators for causal effect estimation in observational studies. Our novel estimators are both consistent and robust to model misspecification, thereby combining the strengths of traditional weighting and subclassification estimators for causal effect estimation from observational studies. Numerical studies show that the proposed estimators perform favorably compared to existing methods.

\vspace{9pt}
\par
\end{quotation}\par

\def\thefigure{\arabic{figure}}
\def\thetable{\arabic{table}}

\renewcommand{\theequation}{\thesection.\arabic{equation}}

\fontsize{12}{14pt plus.8pt minus .6pt}\selectfont

\setcounter{section}{1} %***
\setcounter{equation}{0} %-1

\noindent {\bf 1  Introduction}

The propensity score, defined as the probability of assignment to a particular treatment conditioning on observed covariates, plays a central role in obtaining causal effect estimates in a variety of settings. In a seminal paper, \cite{rosenbaum1983central} showed that adjusting for the propensity score would be sufficient for removing confounding bias due to observed covariates. Since then, many propensity score adjustment methods have been proposed for causal effect estimation. One popular approach is propensity score weighting, which reweights the individuals within each treatment group to create a pseudo-population in which the treatment is no longer associated with the observed covariates. Examples include inverse probability weighting (IPW) estimators \citep[e.g.,][]{rosenbaum1987model,robins2000marginal,liu2016inverse},  doubly robust (DR) estimators \citep[e.g.,][]{robins1994estimation}, and non-parametric weighting estimators \citep[e.g.,][]{hirano2003efficient}. Much of the popularity of weighting estimators comes from their theoretical appeal. For instance, under correct specification of the propensity score model, one can show that the IPW estimators and the classical DR estimator are all consistent for estimating the average treatment effect (ATE). The latter  attains the semiparametric efficiency bound if the analyst  correctly specifies an additional outcome regression model. 

The major criticism of weighting methods is that they are sensitive to misspecification of the propensity score model \citep[e.g.][]{kang2007demystifying}.   Although  in principle, non-parametric models can be used for the propensity score \citep[e.g.][]{hirano2003efficient,mccaffrey2004propensity}, they may not perform well in practice due to the curse of dimensionality. Over the past decade, there have been many endeavors to make the weighting estimators more stable and robust to model misspecification, especially for the classical DR estimator proposed by \cite{robins1994estimation}. Most of these methods construct robust weights by deliberately incorporating the outcome data; see \cite{rotnitkzy:vansteelandt:2014} for a review. In a more recent stream of literature, researchers instead construct robust \ps weights
by balancing empirical covariate moments \citep{hainmueller2011entropy,imai2014covariate,zubizarreta2015stable,chan2016globally,wong2017kernel}. In contrast to the previous approach, the covariate-balancing  \ps weights are constructed without using the outcome data. This leads to two advantages. Firstly, as advocated by \cite{rubin2007design}, the design stage, including the analysis of data on the treatment assignment, should be conducted prior to seeing any outcome data. The separation of the design stage from the analysis stage   helps prevent selecting models that favor ``publishable'' results, thereby ensuring the objectivity of the design.  Secondly, this separation widens the applicability of a robust weighting scheme as, in principle, it could be applied to any weighting estimators \citep{imai2014covariate}.

In this paper, we propose an alternative approach to construct robust \ps weights, by borrowing ideas from the subclassification estimator \citep{rosenbaum1984reducing}.
Rather than adjusting weights to try to achieve covariate balance, we aim directly for  robust \ps weights by employing a rank-based approach at the design stage.
Specifically, we initially fit a parametric propensity score model, but we use this solely to sort  and cluster the units; the propensity weights are then computed as empirical averages within each group. Similar to the covariate-balancing \ps weights, our subclassification weights are constructed independently of the outcome data. Furthermore, our approach enjoys several other attractive properties. First, under correct propensity score model specification (and the positivity assumption), the ATE is estimated consistently,  regardless of the response pattern. Second, relative to weights estimated parametrically, there is a dramatic improvement in both weight stability and covariate balance, especially when the propensity score model is misspecified. Third, with the proposed weights, different weighting estimators tend to give similar answers; in particular, two popular IPW estimators coincide with each other (see Proposition \ref{prop:coincide}).
As we discuss later in Section \ref{subsec:cbw}, none of the existing covariate-balancing methods has all these properties.

Our proposal is also closely related to the propensity score subclassification estimator for the ATE. 
The conventional subclassification estimator is constructed by grouping units into several subclasses based on their estimated propensity scores, so that the propensity scores are {approximately} balanced in all treatment groups within each subclass \citep{rosenbaum1984reducing}. Analysts then take weighted averages of the crude effect estimates within each subclass to estimate the ATE. The convention is to subclassify at  \emph{quintiles} of estimated propensity scores (even for substantial sample sizes), since it has been claimed that this will remove over 90\% of the confounding bias due to observed covariates \citep{rosenbaum1984reducing}. 
However, due to residual confounding, the conventional  subclassification estimator is inconsistent for estimating the ATE.
In contrast, when combined with the Horvitz-Thompson estimator or the classical DR estimator, our proposed \ps weighting scheme leads to novel estimators that are (root-N) consistent.

%The \fss method can be used for robust estimation of propensity score weights, so that the resulting weighting estimators are both consistent and robust to model misspecification, thereby enjoying the advantages of both the classical weighting and subclassification methods.

% Despite suffering from residual confounding, propensity score subclassification methods have been popular among some practitioners because they are relatively robust  to model misspecification  \citep{drake1993effects,kang2007demystifying} and likely to have smaller variance in large samples  \citep{williamson2012variance}. 
%
%In this paper, we instead propose to use subclassification for estimating the propensity score weights. Our approach is based on the fact that the subclassification estimator can be seen as an IPW estimator with weights coarsened via subclassification \citep[Section 3.1; see also][\S 17.8]{imbens2015causal}. On the other hand, the IPW estimators can be seen as the limit of  subclassification estimators as the number of subclasses goes to infinity \citep{rubin2001using}. The main difficulty in formalizing this intuition, however, is that due to 
%

% fail to achieve one of more of these 
%
%
%  To the best of our knowledge, the \fss weighting method is the only one that achieves all these properties. We refer interested readers to Section \ref{subsec:cbw}
%   for a detailed discussion of related methods.

The rest of this article is organized as follows. In Section \ref{sec:background}, we give a brief overview of relevant propensity score 
adjustment methods. In Section \ref{sec:methodology}, we introduce the subclassification weighting scheme and discuss its theoretical properties. We also relate our approach to covariate balancing weighting schemes in the literature, and discuss further beneficial properties of our method. Sections \ref{sec:simulation} and \ref{sec:nhanes} contain
simulations and an illustrative data analysis. 
We end with a discussion in Section \ref{sec:discussion}.

\par
\bigskip

\section{Background}
\label{sec:background}

\subsection{The propensity score}
Let $Z$ be the treatment indicator (1=active treatment, 0=control) and  $\bm X$ denote baseline covariates.  We assume each subject has two potential outcomes $Y(1)$ and $Y(0)$, defined as the outcomes that would have been observed had the subject received the active treatment and control, respectively. We make the consistency assumption such that the observed outcome $Y$ satisfies
$
Y = Z Y(1) + (1-Z) Y(0).
$
We also assume that there is no interference between study subjects  and there is only one version of treatment \citep{rubin1980comment}.
Suppose that we independently sample $N$ units from the joint distribution of $(Z,\bm X,Y)$, and denote them as 
triples $(Z_i, \bm X_i, Y_i), i=1,\ldots,N$.

\cite{rosenbaum1983central} introduced the propensity score $\pi(\bm X)=pr(Z=1\mid \bm X)$ as the probability of receiving the active treatment conditioning on observed covariates. They 
showed that adjusting for the propensity score was sufficient for removing confounding bias under the following assumption:
\begin{assumption}
	\label{assump:strong_igno}
	Strong Ignorability of Treatment Assignment: the treatment assignment is uninformative of the potential outcomes given observed covariates. Formally, $Z \ind \{Y(0),Y(1)\} \mid  \bm X$.
\end{assumption}
It is worth noting that while the covariates may be high-dimensional, the propensity score is always one-dimensional and lies within the unit interval. This dimension reduction property of  the \ps is partly responsible for its popularity.

%The key assumption for identifying $\Delta$ from an observational study is the  ignorable treatment assignment assumption \citep{rosenbaum1983central}, which we maintain throughout this paper:
%
%\begin{remark}
%	Although  results in this paper only rely on the weaker assumption that 
%	$
%		Z \ind Y(z) \mid \bm X, z=0,1,
%	$
%	we keep Assumption \ref{assump:strong_igno} to follow convention.
%\end{remark}	

\subsection{Propensity score weighting estimators}
\label{sec:weighting}

Propensity score is often used to obtain unbiased causal effect estimates via weighting. As a leading example, consider the estimation problem of ATE, namely
$
\Delta = E\{Y(1)\} - E\{Y(0)\},
$
where $E\{\cdot\}$ denotes expectation in the population.

In its simplest form, the IPW estimator for $\Delta$ is known as the \HT estimator and weights individual observations by the reciprocal of the estimated propensity scores \citep{horvitz1952generalization,rosenbaum1987model}:
\begin{equation*}
\hat{\Delta}_{HT}  =  \dfrac{1}{N} \sum\limits_{i=1}^N \dfrac{Z_i Y_i}{\hat{\pi}_i} - \dfrac{1}{N} \sum\limits_{i=1}^N \dfrac{(1-Z_i) Y_i}{1-\hat{\pi}_i},
\end{equation*}
where $\hat{\pi}(\bm X)$ is the estimated propensity score and $\hat{\pi}_i = \hat{\pi}(\bm X_i)$. There have been many estimators which are based on refinements of $\hat{\Delta}_{HT}$, including the Ratio estimator which normalizes the weights in the \HT estimator within the active treatment and control group \citep{hajek1971comment}:
\begin{equation*}
\hat{\Delta}_{Ratio} =  \dfrac{\sum\limits_{i=1}^N Z_i Y_i/\hat{\pi}_i}{\sum\limits_{i=1}^N Z_i /\hat{\pi}_i} - \dfrac{\sum\limits_{i=1}^N (1-Z_i) Y_i/(1-\hat{\pi}_i)}{\sum\limits_{i=1}^N (1-Z_i) /(1-\hat{\pi}_i)},
\end{equation*}
and the classical doubly robust estimator \citep{robins1994estimation}:
\begin{equation}
\label{eqn:dr}
\hat{\Delta}_{DR} = \dfrac{1}{N} \sum\limits_{i=1}^N \dfrac{Z_i Y_i - (Z_i-\hat{\pi}_i) \hat{b}_1(\bm X_i)}{\hat{\pi}_i} - \dfrac{1}{N} \sum\limits_{i=1}^N \dfrac{(1-Z_i) Y_i + (Z_i-\hat{\pi}_i)\hat{b}_0(\bm X_i)}{1-\hat{\pi}_i},
\end{equation}
where $\hat{b}_z(\bm X)$ is an estimate of $E(Y|Z=z,\bm X)$ obtained via outcome regression. 

The propensity score weighting estimators are  attractive theoretically. For example, under correct model specification, $\hat{\Delta}_{DR}$ attains the semiparametric efficiency bound; furthermore, it is doubly robust in the sense that it is consistent if either the propensity score model or the outcome regression model is correctly specified. However, this theoretical appeal may be diminished in practice because of sensitivity to model misspecification \citep[e.g.][]{kang2007demystifying}. 
%Instead, the subclassification estimators are more robust to model misspecification \citep{drake1993effects,kang2007demystifying}.

Besides the ATE, alternative estimands that are popular in practice include the  multiplicative causal effect  $E\{Y(1)\} / E\{Y(0)\}$, and the average treatment effect on the treated $E\{Y(1)-Y(0)\mid Z=1\}.$ Although we use  $\Delta$ as an example, our focus in this paper is on estimation of propensity score weights, and  methodologies introduced here are applicable to these alternative estimands.

\subsection{Propensity score subclassification estimators for the ATE}
\label{subsec:subclassification}

The propensity score subclassification estimator for the ATE
involves stratifying  units into subclasses based on estimated propensity scores, and then directly comparing treated and control units within the same subclass \citep{rosenbaum1984reducing}. Formally, let  $[\hat{\pi}_{min},\hat{\pi}_{max}]$ be the range of estimated propensity scores;
$\hat{C}_k = [\hat{q}_{k-1},\hat{q}_k)\ (k=1,\ldots,K)$ be disjoint divisions of the interval $[\hat{\pi}_{min},\hat{\pi}_{max})$; ${n_k} = \sum_{i=1}^N I(\hat{\pi}_i \in \hat{C}_k)$ and $n_{zk} = \sum_{i=1}^N I(\hat{\pi}_i \in \hat{C}_k)I(Z_i=z), z=0,1$.
Then the subclassification estimator is
\begin{equation*}
\label{eqn:ATE_S}
\hat{\Delta}_S =  \sum\limits_{k=1}^K \dfrac{{n_k}}{N}  \left\{   \dfrac{1}{n_{1k}} \sum\limits_{i=1}^N Z_iY_i I(\hat{\pi}_i\in \hat{C}_k)     - \dfrac{1}{n_{0k}} \sum\limits_{i=1}^N (1-Z_i)Y_i I(\hat{\pi}_i\in \hat{C}_k) \right\}.
\end{equation*}
Note that due to strong ignorability of the propensity score, we have
\begin{equation}
\label{eqn:strong_igno}
\Delta = E_{\pi(\bm X)}\left[E\{Y|\pi(\bm X),Z=1\} - E\{Y|\pi(\bm X),Z=0\} \right].
\end{equation}
The subclassification estimator can hence be viewed as a histogram approximation to \eqref{eqn:strong_igno}.

Most applied publications choose $K$ to be $5$ based on \cite{rosenbaum1984reducing}'s recommendation, in which case the cut-off points are often chosen as sample quintiles.
It is well-known that when $K$ is fixed,  $\hat{\Delta}_S$ is biased and inconsistent for estimating $\Delta$ due to residual bias \citep[e.g., ][]{lunceford2004stratification}.

%\begin{itemize}
%	\item Horvitz-Thompson IPW estimator
%	\item Ratio estimator
%	\item Doubly-robust estimator
%\end{itemize}

\section{Methodology}
\label{sec:methodology}

In this section, we describe a novel robust method to estimate propensity score weights. 
Typically in practice, researchers assume a parametric model for the propensity score, i.e.
$$
pr(Z=1\mid \bm X) =	\pi(\bm X;\beta).
$$
However, it is well-known that even under mild model misspecification, parametric model-based propensity score estimates may be very close to 0 or 1, causing their reciprocals to behave wildly. To address this, we develop subclassification-based propensity score weights to mitigate the bias due to misspecification of the parametric models. We achieve this robustness by using only the rank information from  model-based propensity score estimates, rather than their specific values.  In what follows, we first outline the subclassification weights, and then discuss the choice of the number of subclasses. We also compare the subclassification weights to various previous proposals for robust estimation of propensity score weights.

\subsection{The subclassification weights}
\label{sec:method}

The subclassification weights can be constructed in the following steps: (i) use model-based propensity score estimates to form subclasses $\hat{C}_k, k=1,\ldots, K$, as in the construction of $\hat{\Delta}_S$; (ii) estimate the subclass-specific propensity score $p_k = pr(Z=1\mid \hat{\pi}_i \in \hat{C}_k)$ by its empirical estimate, i.e. $n_{1k}/n_k$; here $p_k$ is well-defined as $\hat{\pi}_1,\ldots,\hat{\pi}_N$ are identically distributed; (iii) use $\hat{p}_k$ as the propensity score estimate for all subjects in subclass $k$. Mathematically, the subclassification weights are defined as 
%\begin{equation}
%\label{eqn:stable_weights}
%w_{S} = \left\{  \begin{array}{ll}
%1/ {\hat{p}_{\hat{k}_i}}  &  \quad \text{if } Z=1\\
%1/ (1-\hat{p}_{\hat{k}_i}) & \quad  \text{if } Z=0\\
%\end{array}  \right.,
%\end{equation}
\begin{equation*}
w_{S} = \frac{Z}{{\hat{p}_{\hat{k}_i}}}
 +\frac{1-Z}{1-\hat{p}_{\hat{k}_i}},
\end{equation*}
where $\hat{k}_i=k$ if $\hat{\pi}_i\in \hat{C}_k.$ Intuitively, the subclassification weights can be viewed as a coarsened version of the original model-based propensity score weights. 

When applied to the Horvitz-Thompson estimator for estimating the ATE, the subclassification weights give rise to the usual subclassification estimator $\hat{\Delta}_S.$ In other words, let superscript $S$  denote the subclassification weighting scheme, we have the following identity:
\begin{equation}
\label{eqn:key}
\hat{\Delta}_S = \hat{\Delta}_{HT}^S = \dfrac{1}{N} \sum\limits_{i=1}^N \left\{ \dfrac{Y_i Z_i}{\hat{p}_{\hat{k}_i}} - \dfrac{Y_i (1-Z_i)}{1-\hat{p}_{\hat{k}_i}} \right\};
\end{equation}
see also \citet[\S 17.8]{imbens2015causal}.
Identity \eqref{eqn:key} motivates the name for subclassification weights. We emphasize that since the subclassification weights are constructed independently of the  outcome data, in principle, it can be applied to any propensity score weighting estimator.

As advocated by \cite{rubin2007design}, the propensity scores should be estimated in a way such that different model-based adjustments tend to give similar answers. In Proposition \ref{prop:coincide}, we show that when used in combination with the subclassification weights,  the \HT estimator coincides with the Ratio estimator.
This is appealing as the latter  has  better statistical properties in terms of both efficiency and robustness \citep{lunceford2004stratification}, while the former is easier to describe and arguably more widely used in practice. 
%The proof is given  in the Supplementary Material.

\begin{proposition}
	\label{prop:coincide}
	If we estimate the propensity score with  ${\hat{p}_{\hat{k}_i}}$, then	the \HT estimator coincides with the Ratio estimator:  $\hat{\Delta}_{HT}^{S} = \hat{\Delta}_{Ratio}^{S}$. 
\end{proposition}

\subsection{Choice of the number of subclasses}
\label{sec:main theorem}

A practical problem in specifying the subclassification weights is choosing the number of subclasses $K$.  Intuitively a large number of subclasses reduces bias, but potentially leads to higher variance. We hence  consider 
increasing the number of subclasses with sample size such that with large enough sample size the coarsened weights can approximate the individual (model-based) weights to an arbitrary level, while with small sample size the coarsened weights are much more stable than the individual (model-based) weights. 

To make our discussion concrete, we apply subclassification weights to the Horvitz-Thompson estimator, and study the rate at which the number of subclasses should increase with the sample size. Formally, we rewrite the Horvitz-Thompson estimator with subclassification weights as follows:
\begin{equation*}
\label{eqn:ATE_VS}
\hat{\Delta}_{HT}^{S} =  \sum\limits_{k=1}^{K(N)} \dfrac{{n_k}}{N}  \left\{   \dfrac{1}{n_{1k}} \sum\limits_{i=1}^N Z_iY_i I(\hat{\pi}_i\in \hat{C}_k)     - \dfrac{1}{n_{0k}} \sum\limits_{i=1}^N (1-Z_i)Y_i I(\hat{\pi}_i\in \hat{C}_k) \right\},
\end{equation*}
where we write $K=K(N)$ to emphasize that the number of subclasses $K$ is a function of the sample size $N$.  With slight abuse of notation, we define $\hat{C}_k, n_k$ and $n_{zk}$ as in Section \ref{subsec:subclassification}, with $K(N)$ replacing $K$ in the original definitions.
Following convention, we stratify at \emph{quantiles} of estimated propensity scores such that 
$n_1 \approx  \cdots \approx n_{K(N)} \approx N/K(N)$. 

%
%The key to the theoretical justification of $\hat{\Delta}_{HT}^{S}$ is studying the rate at which the number of subclasses should increase with the sample size, to which we now turn.
%
%
%In particular, when we set $K(N) = K_{max}$,  \eqref{eqn:stable_weights} is called  the \emph{full subclassification weight}. Compared with the standard inverse probability weight, the (full) subclassification weight can be viewed as replacing $\hat{\pi}_i$ with the coarsened estimate $\hat{p}_{\hat{k}_i}$.

We now discuss the choice for the number of subclasses in $\hat{\Delta}_{HT}^{S}$. Intuitively, $K(N)$ should increase fast enough with $N$ so that the residual bias is negligible asymptotically. This is formalized in Theorem \ref{thm:ATE_VS_consistent}, with proof in the Supplementary Material.
\begin{theorem}
	\label{thm:ATE_VS_consistent}
	Assume that Assumption \ref{assump:strong_igno} and the regularity conditions in the Supplementary Material hold, $\hat{\Delta}_{HT}^{S}$ is well-defined  and 
	\begin{flalign}
	%    \label{assump:k->infty} K(N) &\rightarrow \infty , \\
	\label{assumption:k>sqrt(N)} K(N) &\rightarrow \infty \text{ as } N \rightarrow \infty.
	\end{flalign}
	Then
	$\hat{\Delta}_{HT}^{S}$ is a consistent estimator for $\Delta$, i.e.
	$	\hat{\Delta}_{HT}^{S} \rightarrow_p  \Delta.$
	If we assume additionally that 
	\begin{equation}
	\label{eqn:lemma2_2}
	K(N)/\sqrt{N} \rightarrow \infty \text{ as } N \rightarrow \infty,
	\end{equation}
	and 
	\begin{equation}
	\label{eqn:cond_k_slow}  \left[K(N)\log\{K(N)\}\right]/{N} \rightarrow 0  \text{ as } N \rightarrow \infty,
	\end{equation}
	then $\hat{\Delta}_{HT}^{S}$ is a root-N consistent estimator for $\Delta$.
\end{theorem}

Recall that the subclassification estimator essentially uses histograms to approximate $\Delta$. 
The key insight given by Theorem \ref{thm:ATE_VS_consistent} is that to achieve the root-N consistency,  smaller bandwidths are needed in the histogram approximation.  This is  similar in spirit to the kernel density estimation methods that use under-smoothing  to achieve the root-N consistency
\citep[e.g.,][]{newey1994series,newey1998undersmoothing,paninski2008undersmoothed}.  

On the other hand, for $\hat{\Delta}_{HT}^{S}$  to be well-defined, 
the number of subclasses should grow slowly enough so that for all subclasses, there is at least one observation from each treatment group. This is formalized in Theorem \ref{lemma: ATE_VS_well}, with proof in  the Supplementary Material.
\begin{theorem}
	\label{lemma: ATE_VS_well}
	Assume that the regularity conditions in  the Supplementary Material hold and \eqref{eqn:cond_k_slow} is satisfied,
% 	\begin{equation}
% 	\label{eqn:cond_k_slow}  \left[K(N)\log\{K(N)\}\right]/{N} \rightarrow 0  \text{ as } N \rightarrow \infty,
% 	\end{equation}
	then $\hat{\Delta}_{HT}^{S}$ is asymptotically well defined: 	$    pr(n_{zk}>0,\text{ for all } z,k) \rightarrow 1.$
\end{theorem}

Theorems \ref{thm:ATE_VS_consistent} and \ref{lemma: ATE_VS_well} provide theoretical guidelines for the choice of $K(N)$.
In practice, we propose to choose the maximum number of subclasses such that $\hat{\Delta}_{HT}^{S}$  is well-defined:
$$
K_{max} = \max\left\{K: \hat{\Delta}_{HT}^{S} \text{ is well-defined}\right\}.
$$
In other words, we choose the largest $K$ such that for all subclasses $\hat{C}_1, \ldots, \hat{C}_K$ there is at least one observation from each treatment group.
The resulting estimator is called the \emph{\fss}estimator:
$$
\hat{\Delta}_{FS} = \sum\limits_{k=1}^{K_{max}} \dfrac{{n_k}}{N}  \left\{   \dfrac{1}{n_{1k}} \sum\limits_{i=1}^N Z_iY_i I(\hat{\pi}_i\in \hat{C}_k)     - \dfrac{1}{n_{0k}} \sum\limits_{i=1}^N (1-Z_i)Y_i I(\hat{\pi}_i\in \hat{C}_k) \right\},
$$
and the resulting subclassification weights are called \emph{the full subclassification weights. }
{It follows from Theorem} \ref{lemma: ATE_VS_well} {that} {$\hat{\Delta}_{FS}$ satisfies the rate conditions in Theorem} \ref{thm:ATE_VS_consistent}; see the Supplementary Material for a detailed argument. 
% \yz{YZ: I feel the proof for this point may have some problems, please see supplementary material S6.}
%It is easy to see that .
The definition of $K_{max}$ does not use information from the outcome data, thus it  is  aligned with the original spirit of propensity score adjustment \citep{rubin2007design}.

%The \fss estimator is closely related to  the full matching 
%	estimator (Rosenbaum, 1991; Hansen, 2004; Stuart, 2010), which creates multiple matched sets
%	such that each matched set contains either one treated subject and more than one control subjects, or one
%	control subject and more than one treated subjects. The full matching estimator is essentially a
%	subclassification estimator with the maximal number of subclasses. Our approach differs from full
%	matching in that we subclassify by quantiles of the observed data, thereby achieving 
%	subclasses with (approximately) equal number of observations. In contrast, the full matching estimator can have different number of units in different subclasses. In addition,  given a parametric propensity score model, the full subclassification estimator
%	is unique, while the optimal full matching estimator depends on the distance measure used for matching.

%Other related estimators of the ATE include the kernel matching estimator \citep{heckman1998matching}, coarsened exact matching \citep{iacus2011multivariate} and regression within subclasses \citep{imbens2015causal}. 

\subsection{{Application to doubly robust estimation}}
\label{sec:dr}

%In this section we apply the proposed full subclassification results to more complex estimators than those in Proposition \ref{prop:coincide}. Specifically we apply the proposed weights to doubly robust estimating procedures producing novel estimators,

In this section, we apply the proposed full subclassification weights to doubly robust procedures, which leads to novel estimators. This illustrates the use of our new weighting scheme beyond the simple estimators discussed in Proposition \ref{prop:coincide}. In particular, we replace the model-based propensity score estimate  $\hat{\pi}_i$ in \eqref{eqn:dr} with our subclassification-based propensity score estimate $\hat{p}_{\hat{k}_i}$:
$$
\hat{\Delta}_{DR}^{S} =  \dfrac{1}{N} \sum\limits_{i=1}^N \dfrac{Z_i Y_i - (Z_i-\hat{p}_{\hat{k}_i}) \hat{b}_1(\bm X_i)}{\hat{p}_{\hat{k}_i}} - \dfrac{1}{N} \sum\limits_{i=1}^N \dfrac{(1-Z_i) Y_i + (Z_i-\hat{p}_{\hat{k}_i}) \hat{b}_0(\bm X_i)}{1-\hat{p}_{\hat{k}_i}}.
$$
The following theorem shows that under the conditions of Theorems \ref{thm:ATE_VS_consistent} and \ref{lemma: ATE_VS_well},   $\hat{\Delta}_{DR}^S$  is doubly robust and locally semiparametric efficient. 

\begin{theorem}
	\label{thm:dr}
	Assume conditions \eqref{assumption:k>sqrt(N)}, \eqref{eqn:cond_k_slow},  the regularity conditions assumed in Theorems \ref{thm:ATE_VS_consistent}, \ref{lemma: ATE_VS_well}, and additional standard regularity conditions hold. If either (i) the model used to estimate $b_z(\bm X_i)$ is correct so that 
	%	\begin{equation}
	%	\label{eqn:parametric}
	$	\Vert \hat{b}_z(\bm X_i) - b_z(\bm X_i)\Vert = O_p(1/\sqrt{N}), z=0,1,$
	%	\end{equation}
	%	$\Vert \hat{b}_z(\bm X_i) - b_z(\bm X_i)\Vert = o_p(1), z=0,1$,
	or (ii) the propensity score model $\pi(\bm X;\beta)$ is correctly specified, then 	
	$\hat{\Delta}_{DR}^S$ is consistent for estimating $\Delta$. If conditions (i) and (ii) hold simultaneously, then 	$\hat{\Delta}_{DR}^S$ is asymptotically linear with an influence function 
	$$
	D_{dr} = \dfrac{Z Y - (Z-\Pi) B_1}{\Pi} -  \dfrac{(1-Z) Y + (Z-\Pi) B_0}{1-\Pi} - \Delta,
	$$
	where $\Pi = \pi(\bm X)$  and $B_z=b_z(\bm X)$. Consequently,
	$$
	    \sqrt{N} (\hat{\Delta}_{DR}^S - \Delta) \rightarrow_d N(0, \sigma_{dr}^2),
	$$
	where $\sigma_{dr}^2 = {\rm Var}(D_{dr})$. 
	 Furthermore,  $\sigma_{dr}^2$ attains the semiparametric variance bound under the nonparametric model. The asymptotic variance $\sigma_{dr}^2$ can be consistently estimated by 
	 $$
	    \hat{\sigma}_{dr}^2 = \dfrac{1}{N} \sum\limits_{i=1}^N \left\{ \dfrac{Z_i Y_i - (Z_i-\hat{p}_{\hat{k}_i}) \hat{b}_1(\bm X_i)}{\hat{p}_{\hat{k}_i}} -  \dfrac{(1-Z_i) Y_i + (Z_i-\hat{p}_{\hat{k}_i}) \hat{b}_0(\bm X_i)}{1-\hat{p}_{\hat{k}_i}} - \hat{\Delta}_{DR}^S\right\}^2.
	 $$
\end{theorem}

%Under a correct parametric outcome regression model, 
%Our condition (i) in  allows for use of flexible machine learning methods for estimation of $b_z(X)$.
% \doubt{In the Supplementary Material we present a proof for Theorem \ref{thm:dr}, and discuss scenarios where condition (i) can be relaxed so that one can use off-the-shelf  nonparametric regression and flexible machine learning tools for estimating the outcome regressions $b_z(\bm X), z=0,1.$} \yz{YZ: I feel the discussion scenario after the proof for Theorem 3 has some problem. In the discussion scenario, $K = C \times {N}^{1/3}$ where $C$ is a constant. However, $K$ can not take this value, because $K$ should satisfy the condition of $K/\sqrt{N}\rightarrow \infty$ as $N\rightarrow \infty$. If you think I am wrong, then 
% based on the following comment by the referee, we need to add some information here: In the discussion of Theorem 3 of the manuscript, it might be better to point out that
% $\|\hat{b}_z(\mathbf{X}_i-b_z(\mathbf{X}_i)\|=o_p(N^{-1/6})$ suffices with a proper choice of $K$, but this sentence is in the supplementary material of the current version.}

In the Supplementary Material we present a proof for Theorem \ref{thm:dr}. From the proof, we can see that if $K = C \times {N}^{1/3}$, where $C$ is a constant, we have 
$$
\Vert\widehat{\Pi} - \Pi \Vert \leq O_p\left(\dfrac{\sqrt{K}}{\sqrt{N}}\right) + O_p\left(\dfrac{1}{K}\right) + O_p\left(\dfrac{1}{\sqrt{N}}\right) = O_p \left(\dfrac{1}{N^{1/3}}\right).$$
 In this case, condition (i) in Theorem 3 can be relaxed so that $	\Vert \hat{b}_z(\bm X_i) - b_z(\bm X_i)\Vert = o_p(1/{N^{1/6}}), z=0,1.$  As this convergence rate is much slower than the parametric convergence rate $O_p(1/\sqrt{N})$, this choice of $K$ allows for  use of off-the-shelf nonparametric regression and flexible machine learning tools for estimating
the outcome regressions $b_z (X),z = 0,1.$ It remains a practical problem, however, to choose the constant $C$ in practice. We leave this as a future topic for investigation.

 \begin{remark}
    In the last two decades, doubly robust estimation
    has been the focus of intensive research and there have been many refinements.  These new estimators often have improved large sample properties and better finite sample performance compared to the original estimator proposed by \cite{robins1994estimation}; see \citet[][\S 2.4]{tan2010bounded} for a list of properties enjoyed by some of these estimators. These developments are orthogonal to our proposal in this paper, which focuses on robust estimation of the propensity score weights rather than improved estimation of the ATE. In fact, a distinctive feature of our novel propensity score weights is that, in principle, they may be combined with these new approaches.  Section \ref{sec:dr} provides an illustration using the doubly robust estimator of \cite{robins1994estimation}.
 \end{remark}

%Condition (i) in Theorem \ref{thm:dr} holds with a correct parametric model for the outcome regressions, in which case the maximum likelihood estimation will ensure that  Moreover,  as we only require consistency of the resulting estimates. Third, if $K(N)$ is chosen according to condition  \eqref{assumption:k>sqrt(N)} but not \eqref{eqn:lemma2_2}, then for Theorem \ref{thm:dr} to hold, one needs to replace condition (i) with the stronger condition \eqref{eqn:parametric}, in which case the outcome regression can only be estimated via parametric modeling.
%if we assume  many examples for b, including parametric/machine learning models; Second, we need root-N consistency, so the rate conditions are crucial;
%

\subsection{{Relation to covariate balancing weighting schemes}}
\label{subsec:cbw}

The \fss weighting method is closely related to covariate balancing weighting methods, which  also aim to achieve robust estimates of the ATE without using the outcome data \citep{hainmueller2011entropy,imai2014covariate,zubizarreta2015stable,chan2016globally,wong2017kernel}. 
These methods 
%are designed to reduce empirical covariate imbalance between the two treatment groups, as it may result in severe bias in the final causal effect estimates. Prior to these methods, practitioners often try multiple propensity score models until a sufficiently balanced solution is found; this cyclic process is known as the propensity score  tautology \citep{imai2008misunderstandings}. To avoid this,  the covariate balancing methods 
directly weight observations in a way that ensures  the empirical moments of pre-specified covariates are balanced in the weighted sample. 
However, as pointed out by \cite{zubizarreta2015stable}, tighter covariate balancing generally comes at the cost of greater weight instability. 
Although the covariate balancing conditions can be used to eliminate biases due to imbalance in moment conditions \citep{hainmueller2011entropy,chan2016globally}, as we show later in empirical studies, eliminating bias may give rise to  extreme weights even with a correct propensity score model.  This instability of weight estimates not only increases the variance of the  final causal effect estimates, but also makes these estimates highly sensitive to outliers in the outcome data. In contrast, as we show later in simulation studies, the \fss weighting method often achieves a good compromise for this covariate balance-stability trade-off.
In related work, \citet{wong2017kernel} presented a more general method that reduced biases arising from imbalance in general functions living in a reproducing-kernel Hilbert space.

The \fss weighting method has a number of advantageous features which are not possessed by  individual covariate balancing methods. 
First, 
%based on the \fss weighting method, the \HT estimator is consistent for estimating the ATE (with a correct propensity score model). In contrast, even under a linear response pattern, SBW only yields an approximately unbiased  estimate of the ATE. 
%This bias remains even with an infinite sample size, yielding the resulting estimator to be inconsistent. 
%Second, 
the reciprocal of the \fss weights have the interpretation of coarsened propensity scores; in particular, they always lie within the unit interval.
Consequently, the \fss weighting methods can be conceptualized as creating a pseudo population through inverse probability weighting.
In contrast,  although  the reciprocal of normalized empirical balancing (EB) weights \citep{hainmueller2011entropy} and empirical balancing calibration (CAL) weights \citep{chan2016globally} imply propensity scores asymptotically, they can be greater than 1 or even negative in small sample settings. This is concerning for many practitioners given the ``black box'' involved in estimating these weights.  Second, our method may use any parametric form for the posited propensity score model, whereas the default versions of the EB and CAL method both implicitly assume a logistic regression model. Third, calculating the \fss weight is a convex problem as long as the parameter estimation in the propensity score model is convex. In contrast,  
it was reported in the literature that even with a logistic regression model, the optimization problem of covariate balancing propensity score \citep[CBPS,][]{imai2014covariate} may be non-convex, so  it may be difficult to find the global optimal solution in practice \citep{zhao2017entropy}.

% Furthermore, at least for practitioners,  it is difficult to generalize these methods to other forms of propensity score models. Fourth, the \fss weight is extremely easy to calculate in practice. Applied researchers can use standard off-the-shelve softwares to fit a familiar propensity score model. 

%   Implementation of this is also highly computationally efficient as no complex optimization problem is involved.
%Second,   These features make the \fsw weight an appealing alternative in practice.

%
%Third, we do not use (moment based) covariate balancing conditions in our construction of robust propensity score weights. As propensity score balance the distribution of covariates, and hence moments of covariates in the treatment groups, 
%practitioners can check the validity of the propensity score by empirically checking the (moment based) covariate balancing conditions. This is similar in spirit to the specification test for overidentified CBPS \citep{imai2014covariate}. In other words, (moment based)  covariate balancing conditions \emph{should be used as diagnostics for the propensity score model, rather than a goal of analysis.} In particular, weighting methods that balance certain moments of observed covariates but do not balance the propensity score  distribution may still yield biased estimates. 

%
%propensity score tautology. it is not meaningful to check covariate balancing conditions if the procedure directly targets at this, in a way that is separated from estimating the propensity score.

\section{Simulation Studies}
\label{sec:simulation}

In this section we evaluate the finite sample performance of the proposed full subclassification  weighting scheme. 
%We compare it to  classical subclassification and weighting estimators, as well as various covariate balancing weighting schemes. 
Our simulation setting is the same as that in 
\cite{chan2016globally}. Under this setting, we also evaluate the variance estimator $\hat{\sigma}^2_{dr}$ described in Theorem \ref{thm:dr}.
%We then simulate data under an alternative setting to  evaluate the sensitivity of simulation results to the shapes of the propensity score and response surface.
%However, as we explain in detail later, the shape of the propensity score model and the outcome regression model in \cite{kang2007demystifying}'s setting happens to satisfy the \emph{extra} implicit assumptions of many covariate balancing methods. We hence modify \cite{kang2007demystifying}'s settings such that the coincidence breaks. This yields a more fair comparison as in the design stage, the shape of the response surface is unknown.

Specifically, 	our simulation data consist of $N$ independent samples from the joint distribution of $(Z,\bm X,Y,\bm W)$.  The covariates $\bm X = (X_1,X_2,X_3,X_4)^\top$ follow a standard multivariate normal distribution $N(0,\mathbf{I}_4)$, where $\mathbf{I}_4$ is a $4\times 4$ identity matrix. The treatment variable $Z$ follows a Bernoulli distribution with mean $\text{expit}(\bm X^\top \bm \gamma)$, where $\bm\gamma = (-1,0.5,-0.25,-0.1)^{\top}$. 
%Although in this setting Assumption \ref{assump:positivity} is violated as the support of the  propensity score distribution is $[0,1]$, we choose  this setting nevertheless to facilitate comparison with existing results. 
Conditional on $\bm X$, the potential outcome {$Y(1)$ is defined by the linear model $Y(1) = 210 + b(\bm X)+ \epsilon$ and $Y(0)$ is defined by the linear model $Y(0) = 200 -0.5 b(\bm X)+ \epsilon$}, where $b(\bm X)=27.4 X_1 +13.7 X_2 + 13.7 X_3 + 13.7 X_4$ and the independent error term $\epsilon$ follows a standard normal distribution.  The observed outcome is generated following the consistency assumption: $Y = Y(z)$ if $Z=z$.  Following \cite{kang2007demystifying}, we consider four combinations based on whether the propensity score  model or the outcome regression model is correctly specified. To correctly specify  the propensity score model, we (correctly) include $\bm X$ in the posited logistic  regression model. Otherwise we include covariates $\bm W=(W_1,W_2,W_3,W_4)^\top$, which are  non-linear transformations of $\bm X$ with $W_1 = \exp(X_1/2), {W_2= X_2/\{1+\exp(X_1)\}}, W_3 = (X_1 X_3/25+0.6)^3$, and $W_4 = (X_2+X_4+20)^2$. The specifications of  the outcome regression model are similar.  We are interested in estimating the  ATE, whose true value is {$10$}.	All the simulation results are based on 1000 Monte Carlo samples.

We first compare the full subclassification estimator $\hat{\Delta}_{FS}$ with the classical subclassification estimator $\hat{\Delta}_S$ (with $K=5$) and the Ratio estimator $\hajek$.  $\hat{\Delta}_{HT}$  is not included as it performs uniformly worse than $\hajek$, and  $\hat{\Delta}_{DR}$  is included later as its performance depends on an additional outcome regression model. 
{For comparison, we include the matching estimator of} \cite{abadie2016matching}, {which is implemented with the default options in {\tt R} package {\tt Matching}. }
%As pointed out by \cite{stuart2010matching}, these  estimators represent a continuum in terms of the number of subclasses formed.  
%Figure \ref{fig:simu_fs} and 
Results are summarized in Table \ref{tab:simu_fs}; {see also Figure S1 in the Supplementary Material}.
%\yz{YZ: I put both figure and table here for easy of comparison, later we can decide which one to use.}
When the propensity score model is correctly specified, the classical subclassification estimator   is not consistent; in fact, its bias stabilizes with increasing sample size. The rest of the estimators are consistent for the ATE. 
%\doubt{Among them, $\hat{\Delta}_{FS}$ has the smallest RMSE, with comparable performance only when the sample size is very small.} \yz{YZ: Now it changes to: when the sample size is less than $2000$, $\hat{\Delta}_S$ has the smallest RMSE; when the sample size is larger
%  than $2000$, $\hat{\Delta}_{FS}$ has the smallest RMSE. }
{The RMSE of $\hat{\Delta}_{FS}$ is smaller than those of $\hat{\Delta}_{Ratio}$ and $\hat{\Delta}_{Match}$.}
% This shows that $\hat{\Delta}_{FS}$ achieves a good balance for the bias-variance trade-off discussed in Section \ref{sec:method}.  
When the propensity score model is misspecified, the Ratio estimator  is severely biased; in fact, its bias and RMSE grow with the sample size. Consistent with previous findings in the literature, the other estimators  are more robust to model misspecification, 
%\doubt{with  $\hat{\Delta}_{FS}$ and the matching estimator exhibiting better performance than $\hat{\Delta}_S$ in term of both bias and RMSE.} \yz{YZ: Now it changes to:}
{with  $\hat{\Delta}_{FS}$ and the matching estimator having smaller bias compared to $\hat{\Delta}_S$.}

%\begin{itemize}
%	\item[{\tt bth}:] both the outcome regression model and the propensity score model are correctly specified;
%	\item[{\tt psc}:]  the propensity score model is correctly specified but the outcome regression model is not;
%	\item[{\tt lop}:] the the outcome regression model  is correctly specified but the propensity score model  is not;
%	\item[{\tt bad}:] both the outcome regression model and the propensity score model  are misspecified.
%\end{itemize}

%
%The number of subclasses was determined by the formula
%\begin{equation}
%	\label{eqn:K(N)}
%	K(N)=nint\left( K_0 \cdot \left(\dfrac{N}{100}\right)^{\alpha} \right),
%\end{equation} where $nint$ is the nearest integer function and
%%$C$ was chosen to be $K_0\cdot 100^{\alpha}$ such that when sample size is 100, the smallest one we considered in our simulation, number of subclasses are the same for all choice of $\alpha$.
%$K_0$ was chosen to be 5 to facilitate comparison with the results in \cite{lunceford2004stratification}.

%
%We ranged $\alpha$ from 0 to 0.8 in our simulation. The case where $\alpha=0$  corresponds to the fixed subclass proposal described in \cite{lunceford2004stratification}. For $\alpha \in (0,0.8]$, our theory predicts consistency of $\Delta_{VS}$.

\begin{table}[ht]
\centering
\caption{Bias and root mean squared error (RMSE) of the classical subclassification estimator (S), the full subclassification estimator (FS), the Ratio estimator (Ratio) and the matching estimator (Match) under  \cite{chan2016globally}'s setting. }
\label{tab:simu_fs}
		
\bigskip
\begin{tabular}{rrrrrrrrr}
  \toprule 
  			Sample size  &       \multicolumn{4}{c}	{ PS model correct} &  \multicolumn{4}{c}	{ PS model incorrect} \\[3pt]
  			
 & S & FS & Ratio & Match & S & FS & Ratio & Match \\ 
  \midrule
  Bias&&&&&&&&\\[5pt]
100 & -0.76 & -0.77 & -0.51 & -0.54 & -1.12 & -0.81 & 1.33 & -0.90 \\ 
  200 & -0.61 & -0.37 & -0.18 & -0.29 & -0.88 & -0.31 & 3.77 & -0.39 \\ 
  500 & -0.60 & -0.24 & -0.05 & -0.23 & -0.97 & -0.31 & 5.42 & -0.39 \\ 
  1000 & -0.35 & 0.17 & 0.17 & 0.15 & -0.77 & 0.02 & 6.62 & 0.07 \\ 
  2000 & -0.54 & -0.02 & -0.04 & 0.00 & -0.92 & -0.10 & 7.93 & -0.09 \\ 
  5000 & -0.54 & -0.02 & -0.02 & -0.02 & -0.91 & -0.07 & 9.84 & -0.08 \\ 
  10000 & -0.58 & -0.05 & -0.06 & -0.04 & -0.95 & -0.10 & 10.12 & -0.08 \\ [1em]
   RMSE&&&&&&&&\\[5pt]
  100 & 6.84 & 6.93 & 7.90 & 7.52 & 6.51 & 6.72 & 9.31 & 7.18 \\ 
  200 & 4.46 & 4.74 & 5.24 & 4.92 & 4.40 & 4.60 & 10.98 & 4.89 \\ 
  500 & 2.78 & 2.97 & 3.39 & 3.10 & 2.82 & 2.81 & 11.33 & 3.06 \\ 
  1000 & 1.93 & 2.09 & 2.41 & 2.23 & 2.03 & 2.01 & 12.12 & 2.13 \\ 
  2000 & 1.43 & 1.43 & 1.69 & 1.54 & 1.61 & 1.38 & 13.63 & 1.51 \\ 
  5000 & 1.01 & 0.92 & 1.04 & 0.99 & 1.23 & 0.87 & 15.73 & 0.97 \\ 
  10000 & 0.83 & 0.65 & 0.75 & 0.69 & 1.12 & 0.63 & 15.56 & 0.67 \\ 
   \bottomrule
\end{tabular}
\end{table}

We then compare various weighting schemes for the three classical weighting estimators introduced in Section \ref{sec:weighting}: $\hat{\Delta}_{HT}, \hat{\Delta}_{Ratio}$, and $\hat{\Delta}_{DR}$. The weights we consider include true weights; logit weights, obtained by inverting propensity score estimates from a logistic regression model; trimmed weights, obtained by trimming logit weights at their {5\% percentiles and 95\% percentiles}; (over-identified) covariate balancing propensity score (CBPS) weights  of \cite{imai2014covariate}; 
%stable weights (SBW) of \cite{zubizarreta2015stable},
empirical balancing calibration weights of \cite{chan2016globally} implied by exponential tilting ({CAL-ET}) or quadratic loss ({CAL-Q}), and  the proposed full subclassification (FS) weights. We use the default options of {\tt R} packages 
{\tt CBPS} and {\tt ATE} for calculating the CBPS and CAL weights, respectively.

\begin{figure}[ht]
	\centering
	\includegraphics[width=\linewidth]{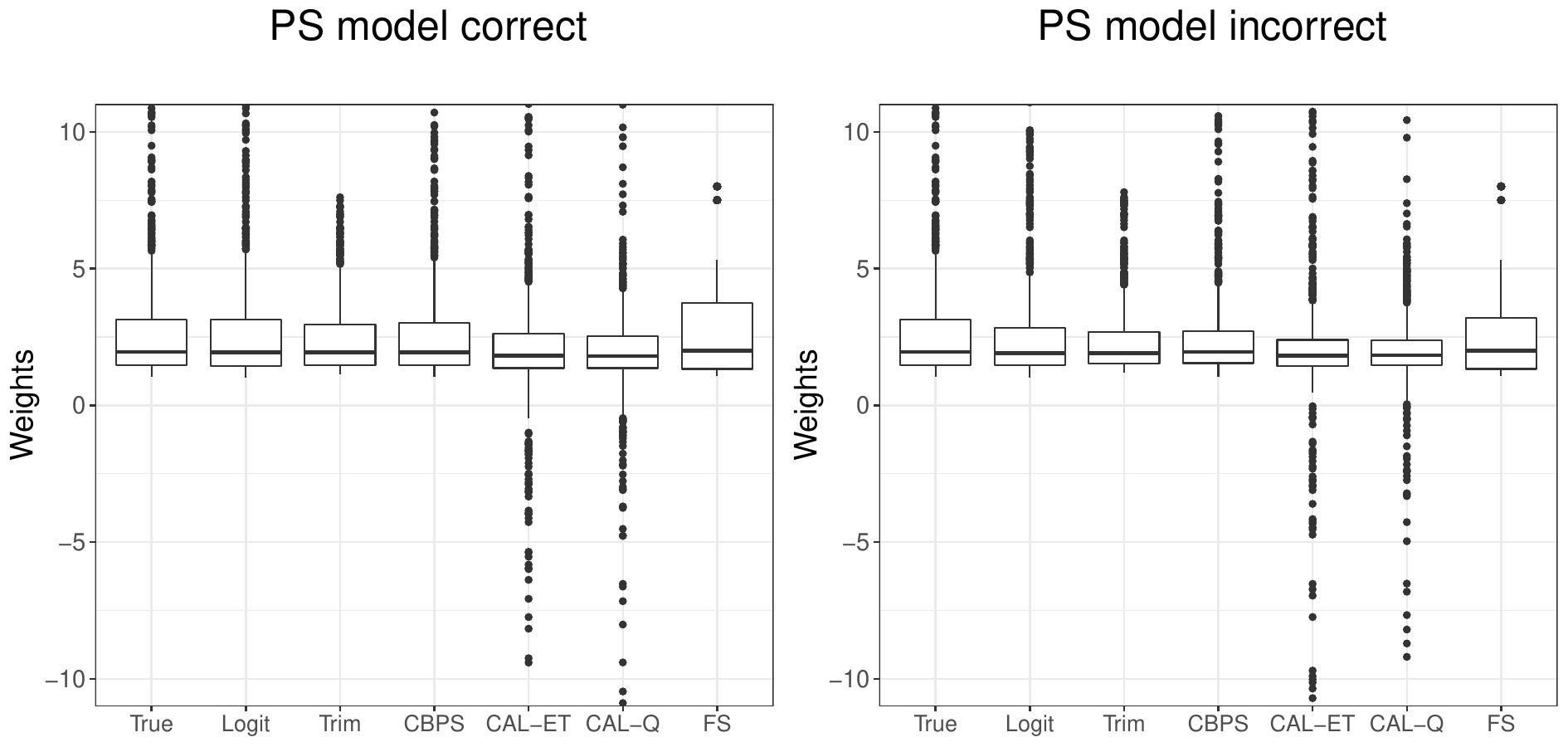}	
	\caption{Distributions of weight estimates with various weighting scheme with a random simulated data set of sample size 1000. Weights outside the range of $[-10,10]$ are truncated;
% 	The figure under the name of weighting scheme shows the percentage (\%) of points whose weight estimates are outside the range of $[-10,10]$.
see also Figure S2 in the Supplementary Material for the full  plot. }
	\label{fig:simu_ks_boxplots2}
\end{figure}

% \begin{table}[ht]
% 	\begin{center}
% 		\caption{Standardized imbalance measures of various weighting schemes under \cite{chan2016globally}'s setting. We consider both 
% 			correct ($\checkmark$) and incorrect ($\times$) specifications of the propensity score (PS) model*.}
% 		\label{tab:imb}
		
% 		\bigskip
% 		\begin{tabular}{rccccccccc}
% 			\toprule
% 			Sample size  &       \multicolumn{1}{c}	{ Model} && \multicolumn{6}{c}	{ Weighting scheme} \\
% 			\cmidrule(r){2-2} \cmidrule(l){4-9}
% 			& \multicolumn{1}{c}{PS} &&   {Logit} &{{Trim}}&  {CBPS} & {CAL-ET} & {CAL-Q}   & {FS}   \\
% 			\midrule
% 			200          &$\checkmark$ && 0.16 & 0.15& 0.19 & 0.00 & 0.00 & 0.16\\
% 			&$\times$     && 0.52 &0.18& 0.19 & 0.00 & 0.00 & 0.17\\
% 			1000         &$\checkmark$ && 0.07 &0.07& 0.09 & 0.00 & 0.00 & 0.07\\
% 			&$\times$     && 0.70 &0.09& 0.11 & 0.00 & 0.00 & 0.08\\
% 			5000         &$\checkmark$ && 0.03 &0.03& 0.04 & 0.00 & 0.00 & 0.03\\
% 			&$\times$     && 6.09 &0.07& 0.12 & 0.00 & 0.00 & 0.06\\
% 			\bottomrule
% 		\end{tabular}
% 	\end{center}
% 	\footnotesize{*: For the covariate balancing weighting schemes, we say the propensity score model is ``correctly specified'' if we impose balancing conditions on $\bm X$, and say the propensity score model is ``misspecified'' if we impose balancing conditions on $\bm W$. 
% 	}
% \end{table}

\begin{table}[ht]
	\begin{center}
		\caption{Standardized imbalance measures of various weighting schemes under \cite{chan2016globally}'s setting. We consider both 
			correct ($\checkmark$) and incorrect ($\times$) specifications of the propensity score (PS) model*.}
		\label{tab:imb}
		
		\bigskip
		\begin{tabular}{rccccccccc}
			\toprule
			Sample size  &       \multicolumn{1}{c}	{ Model} && \multicolumn{6}{c}	{ Weighting scheme} \\
			\cmidrule(r){2-2} \cmidrule(l){4-9}
			& \multicolumn{1}{c}{PS} &&   {Logit} &{{Trim}}&  {CBPS} & {CAL-ET} & {CAL-Q}   & {FS}   \\
			\midrule
			200          &$\checkmark$ && 0.16 & 0.14& 0.19 & 0.00 & 0.00 & 0.16\\
			&$\times$     && 0.52 &0.18& 0.19 & 0.00 & 0.00 & 0.17\\
			1000         &$\checkmark$ && 0.07 &0.06& 0.09 & 0.00 & 0.00 & 0.07\\
			&$\times$     && 0.70 &0.08& 0.11 & 0.00 & 0.00 & 0.08\\
			5000         &$\checkmark$ && 0.03 &0.03& 0.04 & 0.00 & 0.00 & 0.03\\
			&$\times$     && 6.09 &0.05& 0.12 & 0.00 & 0.00 & 0.06\\
			\bottomrule
		\end{tabular}
	\end{center}
	\footnotesize{*: For the covariate balancing weighting schemes, we say the propensity score model is ``correctly specified'' if we impose balancing conditions on $\bm X$, and say the propensity score model is ``misspecified'' if we impose balancing conditions on $\bm W$. 
	}
\end{table}

As part of the design stage, we use Figure \ref{fig:simu_ks_boxplots2} and Figure S2 in the Supplementary Material  to visualize the weight stability of various weighting schemes and Table \ref{tab:imb} to assess the covariate balance after weighting. 
% \yz{Q1: I feel the two boxplots both provide some important information. If  we have enough space, maybe we can put both of them in the main paper. Q2: the proportion of points whose weight estimates are outside of $[-10,10]$ is high for the propose full subclassification method, do we need to adjust the range? } 
The covariate balance  is measured using the standardized imbalance measure \citep{rosenbaum1985constructing}:
\begin{multline*}
\text{Imb} = \Bigg\{\left( \dfrac{1}{N} \sum\limits_{i=1}^N \left[(Z_i w_{1i}  - (1-Z_i) w_{0i})\bm  X_i \right]^\top  \right)
\left(\dfrac{1}{N} \sum\limits_{i=1}^N \bm X_i \bm X_i^\top \right)^{-1}\\
\times
\left( \dfrac{1}{N} \sum\limits_{i=1}^N (Z_i w_{1i} - (1-Z_i)w_{0i})\bm  X_i   \right)
\Bigg\}^{1/2},
\end{multline*}
where $w_{1i}$ are weights for the treated and $w_{0i}$ are weights for the control. 
The logit weights perform reasonably well with a correctly specified propensity score model. However, with the misspecified propensity score model, they become highly unstable and cause severely imbalanced  covariate distributions between the two treatment groups.
The CAL weights may look very appealing  as by design, they   achieve exact balance in the standardized imbalance measure between the two treatment groups.  However, as one can see from Figure \ref{fig:simu_ks_boxplots2} and Figure S2, they are highly unstable even under correct specification of the propensity score model. 
Consequently, the causal effect estimate may be driven by some highly influential observations. Moreover, the CAL weights can even be negative for many units.    
In contrast, the {trimmed,} CBPS  and FS weights improve upon the logit weights in term of both stability and covariate balance, with {trimmed and FS weights exhibiting uniformly better performance than CBPS weights}. The performance of FS on covariate balance is particularly impressive as it does not (directly) target at achieving covariate balance between the two treatment groups.

\begin{table}
	\begin{center}
		\caption{Bias and RMSE of classical weighting estimators with various weighting schemes under \cite{chan2016globally}'s setting. We consider both correct ($\checkmark$) and incorrect ($\times$) specifications of the propensity score (PS) model* or the outcome regression (OR) model. The sample size is 1000.}
		% Report071521.rmd
		\bigskip
		\label{tab:est}
		\begin{tabular}{rcccccccccccccccc}
			\toprule
			Estimator  &       \multicolumn{2}{c}	{ Model} &  \multicolumn{7}{c}	{ Weighting scheme} \\
			\cmidrule(r){2-3} \cmidrule(l){4-10}
			& \multicolumn{1}{c}{PS} & \multicolumn{1}{c}{OR} &   \multicolumn{1}{c}{{True}} & \multicolumn{1}{c}{Logit} & {Trim} &  {CBPS} & {CAL-ET} & {CAL-Q}   & {FS}   \\
			\midrule
			\multicolumn{1}{l}{Bias \quad \quad}	 & & & & &&&& \\	
			$\hat{\Delta}_{HT}$  & $\checkmark$  & $-$  & -0.02 & 0.22 & -0.12 & -0.34 & 0.14 & 0.14 & 0.17 \\ 
			$\hat{\Delta}_{HT}$  & $\times$ & $-$ & $-$ & 38.67 & -4.20 & 9.91 & 0.07 & 1.80 & 0.02 \\ 
			$\hajek$  & $\checkmark$  & $-$ & 0.12 & 0.17 & 0.14 & -0.18 & 0.14 & 0.14 & 0.17 \\  
			$\hajek$  & $\times$ & $-$ & $-$ & 6.62 & -2.50 & 1.25 & 0.07 & 1.80 & 0.02 \\ 
			$\hat{\Delta}_{DR}$  & $\checkmark$   & $\checkmark$ &  0.14 & 0.14 & -0.87 & 0.14 & 0.14 & 0.14 & 0.14 \\ 
			$\hat{\Delta}_{DR}$  & $\checkmark$   & $\times$ &0.30 & 0.29 & -0.85 & 0.57 & 1.67 & 2.97 & 0.43 \\ 
			$\hat{\Delta}_{DR}$  & $\times$ & $\checkmark$ & $-$ & 0.14 & -3.18 & 0.14 & 0.14 & 0.14 & 0.14 \\ 
			$\hat{\Delta}_{DR}$  & $\times$ & $\times$ & $-$ &-10.06 & -4.27 & -1.75 & 0.07 & 1.80 & -0.78 \\[12pt]
			\multicolumn{1}{l}{RMSE \quad \quad}	 & & & & &&&& \\
			$\hat{\Delta}_{HT}$  & $\checkmark$  & $-$ & 16.98 & 7.76 & 5.09 & 6.67 & 1.72 & 1.72 & 2.09  \\ 
			$\hat{\Delta}_{HT}$  & $\times$ & $-$ & $-$ & 136.54 & 6.78 & 12.61 & 1.91 & 2.62 & 2.01 \\ 
			$\hajek$  & $\checkmark$  & $-$ & 2.57 & 2.41 & 1.95 & 2.23 & 1.72 & 1.72 & 2.09 \\ 
			$\hajek$  & $\times$ & $-$ & $-$ & 12.12 & 3.26 & 2.53 & 1.91 & 2.62 & 2.01 \\ 
			$\hat{\Delta}_{DR}$  & $\checkmark$   & $\checkmark$ & 1.72 & 1.72 & 1.80 & 1.72 & 1.72 & 1.72 & 1.72 \\ 
			$\hat{\Delta}_{DR}$  & $\checkmark$   & $\times$ & 2.23 & 2.18 & 1.89 & 2.11 & 2.47 & 3.50 & 2.16 \\ 
			$\hat{\Delta}_{DR}$  & $\times$ & $\checkmark$ & $-$ & 1.74 & 3.57 & 1.72 & 1.72 & 1.72 & 1.72 \\  
			$\hat{\Delta}_{DR}$  & $\times$ & $\times$ & $-$ &53.24 & 4.59 & 2.93 & 1.91 & 2.62 & 2.07 \\ 
			\bottomrule
		\end{tabular}
	\end{center}
	\footnotesize{*: For the covariate balancing weighting schemes, we say the propensity score model is ``correctly specified'' if we impose balancing conditions on $\bm X$, and say the propensity score model is ``misspecified'' if we impose balancing conditions on $\bm W$. 
	}
\end{table}

Table \ref{tab:est} summarizes the performance of 
various weighting schemes when they are applied to the three classical weighting estimators. For brevity, we only show results with sample size fixed at 1000. 
Consistent with the findings of \cite{kang2007demystifying}, logit weights are sensitive to misspecification of the propensity score model, regardless of whether the weighting estimator is doubly robust or not.
Use of the full subclassification weights or the covariate balancing  weights ({CBPS}, {CAL-ET} and {CAL-Q}) greatly improves upon the naive weights obtained from a logistic regression model.  
Among them, {FS}, {CAL-ET} and {CAL-Q} weights perform better than {CBPS weights} under {most} simulation settings,
% \yz{(YZ: The {CAL-Q} weight does not perform better than the {CBPS weight}  for  $\hat{\Delta}_{Ratio}$ when the propensity score model is misspecified.  {FS}, {CAL-ET} and {CAL-Q} weights do not perform better than the {CBPS weight}  for  $\hat{\Delta}_{DR}$ when the propensity score model is correctly specified and the outcome regression model is  misspecified.)} 
and the IPW estimator coincides with the Ratio estimator with the {FS}, {CAL-ET} and {CAL-Q} weights.  One noticeable exception is the scenario where the propensity score model is correctly specified while the outcome regression model is misspecified. In this case, when combined with $\hat{\Delta}_{DR},$ CAL-ET and CAL-Q produce estimates with large bias. This is not surprising as neither of these two estimators relies on an explicit propensity score model, so unlike CBPS and FS, their performance does not directly benefit from correct specification of the propensity score model. 
% \doubt{Within these three weights, {CAL-ET} and {CAL-Q} tend to perform better when the propensity score model is correctly specified;  otherwise, {FS} performs better.} 
% \yz{Within these three weights, {CAL-ET} weight tends to perform better than the other two weights in terms of RMSE  except for $\hat{\Delta}_{DR}$ when the propensity score model is correctly specified and the outcome regression model is misspecified. When the propensity score model is correctly specified, {CAL-Q} weight tends to perform better than the FS weight in terms of RMSE except for $\hat{\Delta}_{DR}$ if the outcome regression model is misspecified. When the propensity score model is misspecified, FS weight tends to perform better than the {CAL-Q} weight in terms of RMSE. }
% As argued by many previous researchers  \citep{zubizarreta2015stable,chan2016globally}, it is very likely that the posited models are wrong in practice. Hence the robustness to model misspecification may be worth more attention than performance under correct model specification.
From Table \ref{tab:est}, one can also see that the proposed  {FS weights}  outperform the trimmed weights.
% \yz{YZ: after adjusting the trimmed weights, in some situations, the trimmed weights perform better than the FS weights, for example, when the propensity score model is correctly specified for the Ratio estimator.}
This suggests that the FS weights improve upon the logit weights beyond  stabilizing the extreme weights.

\begin{figure}[!htbp]
    \centering
    \includegraphics[width=\textwidth]{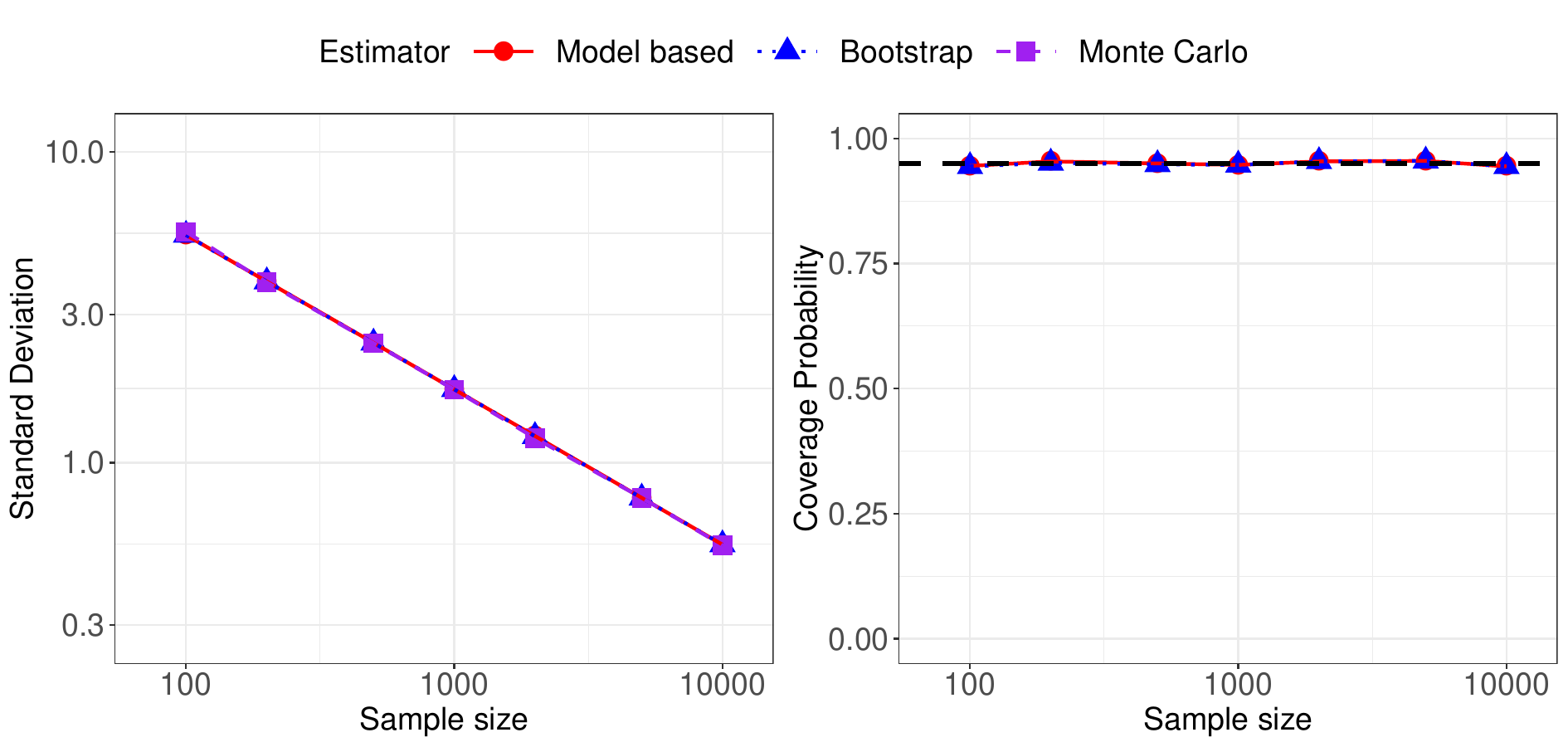}
    \caption{Performance of model-based variance estimator for $\hat{\Delta}_{DR}^S.$ The black dashed line on the right panel shows the 95\% nominal coverage level.}
    \label{fig:my_label}
\end{figure}

Figure \ref{fig:my_label}  presents results on interval estimates. 
 Following Theorem  \ref{thm:dr}, we consider the setting where both conditions (i) and (ii) hold. We compare interval estimates based on two  variance estimators: $\hat{\sigma}_{dr}^2$ described in Theorem  \ref{thm:dr}, and the variance estimator based on 1000 non-parametric bootstrap samples. The left panel in Figure \ref{fig:my_label} shows that estimates from the corresponding standard deviation estimators are close to the Monte Carlo standard deviation, and decrease with the sample size at  approximately the root-N rate. 
The right panel in Figure \ref{fig:my_label} shows the empirical coverage probability of the 95\% Wald-type confidence intervals constructed based on the two variance estimators. Both of these confidence intervals have approximately the nominal coverage level as the sample size increases.

\section{Application to a childhood nutrition study}
\label{sec:nhanes}

We illustrate the application of the proposed full subclassification weighting method using data from the 2007-2008 National Health and Nutrition Examination Survey
(NHANES), which is a program of studies designed to assess the health and nutritional status of adults and children in the United States.  The data set we use was created by \cite{chan2016globally},
which contained  observations on 2330 children aged from 4 to 17. Of these children, 55.1\% participated in the National School Lunch or the School Breakfast programs.
These were federally funded meal programs primarily 
designed to  provide meals for children from poor neighborhoods  in the United States. 
%All children in participating schools are  to receive federally subsidized  meals; in particular, children from families with incomes at or below 130 percent of the poverty level are eligible for free meals.
However, there have been concerns that  meals provided through these programs may cause childhood obesity \citep{stallings2010school}. 
Hence here we study how participation in these meal programs contributes to childhood obesity as   measured by  body mass index (BMI). 
We control for the same set of potential confounders in our analysis as \cite{chan2016globally}; see the Supplementary Material for a detailed list.

%
%	For this survey data, our exposure of interest is self-reported smoking status. For notational convenience, we  define the exposed group to be subjects who never smoked at the baseline visit HALS1. 
%	Our outcome of interest is the change in memory scores, which is considered as a  measure for cognitive functioning. More precisely, the outcome is defined as the difference in memory scores between the baseline  and  follow-up visit.

Table S1 in the Supplementary Material summarizes baseline characteristics and outcome measure by participation status in the school meal programs.  Children participating in the school meal programs are more likely to be black or Hispanic, and come from a family with lower social economic status. Respondents for such children also tend to be younger and female. These differences in baseline characteristics suggest that the observed mean difference in BMI, that is 0.53 $kg/m^2$ {(95\% CI (0.11, 0.96))}, may not be fully attributable to the school meal programs.

\begin{figure}
	\centering
	\includegraphics[width=\linewidth]{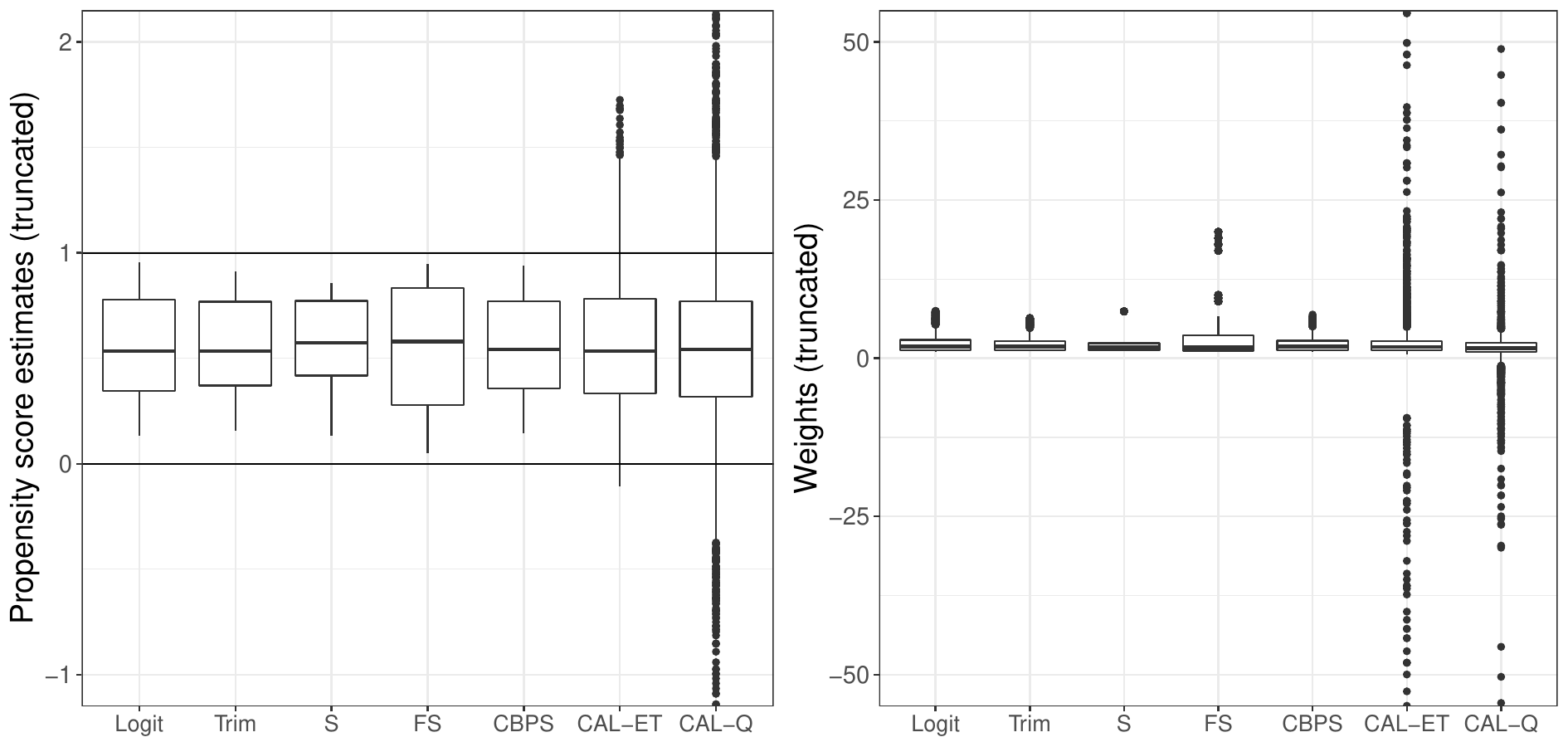}	
	\caption{Distributions of propensity score estimates and weights with the NHANES data.}
	\label{fig:nhanes_boxplots}
\end{figure}

We then apply various weighting  methods to estimate the  effect of participation in the meal programs.
We consider two models for the propensity score: a logistic model and a complementary log-log model.  We also consider a linear outcome regression model on the log-transformed BMI. All the covariates  enter the propensity score model or the outcome regression model as linear terms.
Figure \ref{fig:nhanes_boxplots} visualizes the distributions of propensity score weights and their reciprocals.  Results with the complementary log-log \ps model are similar to those with the logistic regression model and are omitted. We can see that the reciprocals of propensity score weights estimated using the full subclassification method or the CBPS method lie within the unit interval. In contrast, the reciprocals of CAL weights can be negative or greater than 1. Hence these weights cannot be interpreted as propensity scores. Furthermore, consistent with our findings in Figures \ref{fig:simu_ks_boxplots2} and S2, the CAL weights are much more dispersed than  other weights. The five most extreme weights estimated by CAL-ET  are $1350,1259,-959,-943,677$, and those for CAL-Q are $324,153,-97,-97,85.$ 
As these weights are obtained independently of the outcome data, the final causal estimates  are highly sensitive to these outliers.

%\yz{Table \ref{tab:nhanes}}
{Table \ref{tab:nhanes2}}  summarizes the standardized imbalance measure and causal effect estimates. 
%\yz{YZ: Question 1: in Table \ref{tab:nhanes}, we fit a model of $E(Y|Z,X)$ based on all the observations, then estimate $E(Y|Z=1,X)$ and $E(Y|Z=0,X)$. In Table  \ref{tab:nhanes2}, we fit the models of $E(Y|Z=1,X)$ and $E(Y|Z=0,X)$ based on the subsamples where $Z=1$ or $Z=0$ separately, then estimate $E(Y|Z=1,X)$ and $E(Y|Z=0,X)$ based on the fitted coefficients. The only differences in Table \ref{tab:nhanes} and Table \ref{tab:nhanes2} are the estimates and confidence intervals for the doubly robust estimators. I prefer to use Table \ref{tab:nhanes2}, because I have checked that the estimates of slopes in $E(Y|Z=1,X)$ and $E(Y|Z=0,X)$ are not the same. 
% \yz{YZ: do we need to explain the meaning of naive estimator? I think not all the people understand the meaning of this. }
As advocated by \cite{rubin2007design}, the construction of propensity score weights should make the final causal effect estimate  insensitive to the weighting estimator used. However,  the propensity score weights estimated with a parametric model or {the CBPS method} tend to give different answers with different weighting estimators. With these weighting methods, a \HT estimator would suggest that participation in the school meal programs led to a significantly lower BMI. The Ratio estimator and DR estimator instead yield estimates that are much closer to zero.  In contrast, the trimmed weights, the subclassification weights (both the classical ones and the FS weights), and the CAL weights have a consistent implication with different weighting estimators that  participation in school meal programs have negligible effects on the BMI. Moreover,  although different parametric propensity score models may give rise to very different causal effect estimates with a \HT estimator, they yield much closer estimates with a (full) subclassification estimator.  These results show that the subclassification methods are robust against propensity score model misspecification.

\begin{table}[ht]
	\begin{center}
		\caption{The average causal effect estimates associated with participation in the school meal programs. The 95\% Wald-type confidence intervals in brackets are computed based on bootstrap estimates.}
		\label{tab:nhanes2}
		\bigskip
		\begin{tabular}{rccccccccccccccc}
			\toprule
			& Imbalance & HT & Ratio & DR \\ 
			\midrule
Naive* & 1.04 & 0.53 (0.11,0.96) & 0.53 (0.11,0.96) & 0.10 (-0.21,0.42) \\ 
  Logit & 0.10 & -1.52 (-2.46,-0.58) & -0.16 (-0.64,0.32) & 0.08 (-0.38,0.54) \\ 
Trim & 0.06 & -0.01 (-1.09,1.06) & -0.00 (-0.53,0.53) & 0.09 (-0.35,0.52) \\ 
  S & 0.08 & -0.12 (-0.61,0.38) & -0.12 (-0.61,0.38) & -0.02 (-0.48,0.44) \\ 
  FS & 0.12 & -0.20 (-0.75,0.36) & -0.20 (-0.75,0.36) & -0.10 (-0.60,0.40) \\ 
  Cloglog** & 0.15 & -2.26 (-3.67,-0.85) & -0.23 (-0.76,0.30) & 0.14 (-0.35,0.63) \\ 
  Trim $\plus$ Cloglog & 0.13 & 0.70 (-0.48,1.88) & 0.20 (-0.33,0.73) & 0.17 (-0.26,0.60) \\ 
  S $\plus$ Cloglog & 0.16 & -0.05 (-0.55,0.44) & -0.05 (-0.55,0.44) & 0.01 (-0.45,0.47) \\ 
  FS $\plus$ Cloglog & 0.14 & 0.01 (-0.57,0.59) & 0.01 (-0.57,0.59) & 0.08 (-0.43,0.60) \\ 
  CBPS & 0.10 & -1.25 (-2.11,-0.40) & -0.05 (-0.49,0.39) & 0.07 (-0.35,0.50) \\ 
  CAL-ET & 0.00 & -0.05 (-0.48,0.39) & -0.05 (-0.48,0.39) & 0.07 (-0.36,0.51) \\ 
  CAL-Q & 0.00 & -0.02 (-0.45,0.42) & -0.02 (-0.45,0.42) & 0.10 (-0.34,0.53) \\ 
			\bottomrule
		\end{tabular}
	\end{center}
	\footnotesize{*: ``Naive'' corresponds to the crude estimator not adjusting for any confounders; \\ **: Cloglog indicates that the propensity scores are estimated with a complementary log-log model. }
\end{table}

\section{Discussion}
\label{sec:discussion}

Propensity score weighting is among the most popular tools for drawing causal inference in various contexts. 
%To choose among these methods, practitioners often face a bias-variance trade-off as the weighting methods can be consistent while the subclassification methods are more robust to model misspecification. In this article, we connect these two approaches by increasing the number of subclasses in the subclassification estimators. We show that the bias of the propensity score subclassification estimator can be eliminated asymptotically if the number of subclasses increases at a certain rate with sample size. 
%In particular, we propose a novel full subclassification estimator that inherits the advantages of both the classical IPW and subclassification method.
However, with parametric specification of the propensity score model, propensity score weighting estimators tend to be very sensitive to model misspecification. In this paper, we investigate robust estimation of propensity score weights, along the line of research by \cite{hainmueller2011entropy,imai2014covariate,zubizarreta2015stable,chan2016globally}. Specifically, we propose  a novel \fsw scheme that only exploits the rank information from the parametric model-based propensity score estimates, and thus improving upon robustness to model misspecification. 
%Moreover, we show that the full subclassification  method can be used for robust estimation of propensity score weights. 
As discussed in detail by \cite{zubizarreta2015stable}, a covariate balance-stability trade-off is key to constructing robust \ps weights. Through extensive empirical studies, we show that the \fss weighting method achieves a good compromise in this trade-off, and dramatically improves upon model-based propensity score weights in both aspects,  especially when the propensity score model is misspecified.

{Our approach in this paper is based on subclassification by the quantiles of the estimated propensity scores. It is an interesting area for subsequent research to consider alternative ways of forming subclasses and providing theoretical justifications for these alternative approaches. We refer interested readers to} \cite{myers2007optimal} {for an earlier investigation} along these lines.

%In this article,
%we have primarily focused on obtaining a good point estimate for the ATE.
%Although an explicit variance formula is available for the classical subclassification estimator with a fixed number of subclasses, it is likely to be complex in real-life situations and previous researchers have suggested using bootstrap estimates instead \citep{williamson2012variance}. The explicit variance formula for the full subclassification estimator is challenging due to the uncertainty in the number of subclasses. Hence in practice, we recommend using bootstrap or subsampling methods to calculate the standard error and associated confidence intervals. 

%
%The full subclassification method in this article could be applied to address the missing data problem under the missing at random (MAR) assumption \citep[see e.g.][]{rubin1978bayesian,gelman2004applied,kang2007demystifying}. 
%It also 
%   Furthermore, 
{We have focused our illustration on causal estimation from observational studies.} Since the full subclassification weights are constructed independently of the outcome data, it can potentially be applied to improve propensity score estimation in other contexts, such as addressing the missing data problem under the missing at random (MAR) assumption \citep[see e.g.,][]{rubin1978bayesian,gelman2004applied,kang2007demystifying},  causal inference with a marginal structural model \citep{robins2000marginal} and in presence of interference \citep{tchetgen2010causal}.  It can also be extended to obtain robust estimates of the generalized propensity score with multi-arm treatments \citep{imbens2000role,imai2004causal}.

%
%
%In this case, bootstrap can be used in practice to estimate variance and  confidence interval. 
%
%
%All the results here extends directly to estimating 
%The causal effect estimation problem is very similar to that of the 
%
%
%
%
%
%Sub-classified weights in MSM
%
%Sub-classified IPW estimators in presence of interference
%
%
%
%
%The machinery we introduced here can also be used to prove consistency of other subclassification based estimators. For example, one can construct a doubly robust estimator with the cross-classifying procedure introduced in \cite{kang2007demystifying}.
%
%
%
%In this article, we 
%
%We are doing undersmoothing: method introduced in  \cite{newey1998undersmoothing} may be used to get better efficiency results.
%

% We don't discuss propensity score matching as it is controversial... \cite{king2015propensity}

\section*{Acknowledgement}

The authors thank Marco Carone and Peter Gilbert for helpful discussions, and the reviewers and the associate editor for their detailed comments which significantly improved the paper. Wang was supported by NSERC grants RGPIN-2019-07052, DGECR-2019-00453, and RGPAS-2019-00093.  Richardson was supported by ONR grant N00014-15-1-2672. Zhou was supported by CNSF grant 81773546.

	\thispagestyle{empty}
\bibliography{causal}
\bibliographystyle{apalike}

%%%%%%%%%%%%%%%%%%%%%%%%%%%%%%%%%%%%%%%%%%%%%%%%%%%%%%%%%%%%%%%%%%%%%%%%%%%%%%%%%%%%%%%%%%%%%%%%%%%%%%%%%%%%%%%%%%%%%%%%%%%%
%\vskip .65cm
%\noindent
%first author affiliation
%\vskip 2pt
%\noindent
%E-mail: (first author email)
%\vskip 2pt
%
%\noindent
%second author affiliation
%\vskip 2pt
%\noindent
%E-mail: (second author email)
% \vskip .3cm
%\centerline{(Received ???? 20??; accepted ???? 20??)}\par

\clearpage

\centerline{\large\bf SUPPLEMENTARY MATERIALS FOR}
\vspace{2pt}
 \centerline{\large\bf ``ROBUST ESTIMATION OF PROPENSITY SCORE WEIGHTS}
\vspace{2pt}
 \centerline{\large\bf VIA SUBCLASSIFICATION''}
\vspace{.25cm}
\vspace{.4cm} \centerline{$^{1}$Linbo Wang, $^{2}$Yuexia Zhang, $^{3}$Thomas S. Richardson, $^{4}$Xiao-Hua Zhou} \vspace{.4cm} \centerline{\it $^{1,2}$University of Toronto, $^{3}$University of Washington, $^{4}$Peking University
} \vspace{.55cm} 
% \fontsize{9}{11.5pt plus.8pt minus
% .6pt}\selectfont
\par

\setcounter{section}{0}
\setcounter{equation}{0}
\def\theequation{S\arabic{section}.\arabic{equation}}
\def\thesection{S\arabic{section}}
\renewcommand{\theassumption}{S\arabic{assumption}}

% \fontsize{12}{14pt plus.8pt minus .6pt}\selectfont

\section{Proof of Proposition 1}
\label{appendix:proof_prop_coincide}

The proof is straightforward by noting that 
$$
\sum\limits_{i=1}^N Z_i / \hat{p}_{\hat{k}_i} = 	\sum\limits_{i=1}^N n_{\hat{k}_i}  Z_i / n_{1\hat{k}_i} = \sum\limits_{k=1}^K \left\{(n_k/n_{1k})  \sum\limits_{i: \hat{k}_i=k} Z_i\right\} = \sum\limits_{k=1}^K \left\{(n_k/n_{1k} ) n_{1k}\right\} = N,
$$
and similarly
$$
\sum\limits_{i=1}^N (1-Z_i) / (1-\hat{p}_{\hat{k}_i}) = N. 
$$

\section{Regularity conditions for Theorem 1}
\label{sec:assumptions}

We now introduce the regularity conditions that will be used for proving (root-N) consistency of $\hat{\Delta}_{HT}^{S}$.

\begin{assumption}
\label{assump:compact}
(Compact set) The parameter $\Delta$ lies in the interior of a compact set $\mathscr{A}$;  $\beta$ lies in the interior of a compact set $\mathscr{B}$. 
 %$\mathcal{A}$
\end{assumption}

\begin{assumption}
	\label{assump:positivity}
	(Uniform positivity)  (i) The  support of  $\pi(\bm X;\beta)$ can be written as $[\pi_{min},\pi_{max}]$, where $\pi_{min}>0, \pi_{max}<1$, and the quantile distribution of $\pi(\bm X;\beta)$ is Lipschitz continuous. (ii)  There exist constants $c_1$ and $c_2$ such that $0<c_1<\pi(\bm X;\bar{\beta})<c_2<1$ almost surely for all $\bar{\beta}\in \mathscr{B}$.
\end{assumption}

 Assumption \ref{assump:positivity} (i) implies that the cumulative distribution function of $\pi(\bm X; \beta)$ has no flat portions between $[\pi_{min}, \pi_{max}]$, or the quantile distribution of $\pi(\bm X;\beta)$ is continuous on $[0,1]$. Violation of this assumption will cause some subclasses to be always empty, and the subclassification estimator to be ill-defined. This problem may be solved by considering  only non-empty subclasses in constructing the subclassification estimator. For simplicity, we do not get into discussion of this issue here.  
%  Assumption \ref{assump:positivity} (ii) is necessary to ensure some values are bounded and some random variables are bounded in probability in the proof. 

\begin{assumption}
	\label{assump:correct_PS}
	(Uniform consistency of estimated propensity scores)  The propensity score model is correctly specified such that for all $N, \hat{\pi}_i \ (i=1,\ldots,N)$ is uniformly convergent in probability to  $\pi_i \ (i=1,\ldots,N)$ at $\sqrt{N}$ rate, where $\hat{\pi}_i=\hat{\pi}(\bm{X}_i)$ and $\pi_i=\pi(\bm{X}_i;\beta)$. Formally, $\sqrt{N}\max\limits_{1\leq i\leq N} |\hat{\pi}_i-\pi_i| = O_p(1)$. 
\end{assumption}

Under a smooth parametric model, the uniformity part in Assumptions \ref{assump:positivity} and \ref{assump:correct_PS} can usually be inferred from uniform boundedness of the maximum norm of covariates, $|\bm X_i|_\infty$ $(i=1,\ldots,N)$. The latter assumption holds if the support of the covariate $\bm X$ is a bounded set in $\mathbb{R}^p$, where $p$ is the dimension of $\bm X$. This assumption has been widely used in the causal inference literature \citep[for example, see][]{hirano2003efficient}.

As an illustration, suppose the true propensity score model is a logistic regression model: 
$$
\pi(\bm X;\beta) = \text{expit}(\bm X^\top \beta) = \dfrac{\text{exp}(\bm X ^\top\beta)}{1+\text{exp}(\bm X^\top \beta)}.
$$
In this case,  $\pi(\bm X_i;\beta)$ is uniformly bounded away from 0 and 1 if $|\bm X_i|_\infty \ (i=1,\ldots,N)$ are uniformly bounded. At the same time,
by the mean value theorem,
\begin{flalign*}
\sqrt{N} \left| \hat{\pi}(\bm X_i) - \pi(\bm X_i;\beta) \right| % &= |expit(\bm X \hat{\beta}) - expit(\bm X \beta)| \\
&=      \sqrt{N} \left|\text{expit} (\bm X_i^\top \tilde{\beta}) \{1-\text{expit}(\bm X_i^\top \tilde{\beta})\} \bm X_i^\top (\hat{\beta} - \beta)\right|\\
&\leq      \sqrt{N} \Vert \bm X_i \Vert_\infty \Vert \hat{\beta} - \beta \Vert_\infty,
\end{flalign*}
where $\hat{\beta}$ is a consistent estimator of  $\beta$, and $\tilde{\beta}$ lies between $\hat{\beta}$ and $\beta$. Hence $\hat{\pi}(\bm X_i) - \pi(\bm X_i;\beta) \ (i=1,\ldots,N)$ is uniformly convergent in probability to zero at $\sqrt{N}$ rate if   $|\bm X_i|_\infty \ (i=1,\ldots,N)$ are uniformly bounded.

\begin{assumption}
	\label{assump:continuity}
	(Smoothness of propensity score model) $\pi(\bm{X};\beta)$, $\partial \pi(\bm{X};\beta)/\partial \beta$ and $\partial^2 \pi(\bm{X};\beta)/\partial \beta\partial \beta^\top$ are continuous functions of $\beta$.
\end{assumption}

\begin{assumption}
	\label{assump:finitemomentY}
	(Finite second moments)  $E(Y^2)<\infty$ and $E\|\partial \pi(\bm{X};\beta)/\partial \beta\|^2<\infty$, where $\|\cdot\|$ denotes the Euclidean norm.
\end{assumption}

Assumptions \ref{assump:continuity} and \ref{assump:finitemomentY} are standard regularity conditions in the M-estimation theory.

% This assumption can be  satisfied for many different kinds of $Y_i$, including the special cases where
%  $Y_i$ follows the normal distribution, binomial distribution or Poisson distribution.

\begin{assumption}
\label{assump:weakcorr}
(Weak correlation between subclasses) 
$\max\limits_{k,j\in\{1,\ldots,K\},k\neq j}\big|\text{cov}(n_{1k},n_{1j}\mid I(\hat{\pi}_i\in \hat{C}_\ell),i=1,\ldots,N,\ell=1,\ldots,K)\big|=O_p(N/K^2)$.
%$E(n_{1k}n_{1j})-E(n_{1k})E(n_{1j})=O(N/(K^2))$.
\end{assumption}

If one employs a sampling splitting procedure in which separate samples are used for constructing the subclasses, and estimating the average treatment effect, then $n_{1k} \ind n_{1j} \mid I(\hat{\pi}_i\in \hat{C}_\ell),i=1,\ldots,N,\ell=1,\ldots,K$. In this case,  Assumption \ref{assump:weakcorr} holds trivially.

\section{Proof of Theorem 1}
\label{sec:proof_VS_consistent}

Under Assumptions S1--S5, it can be proved using the standard M-estimation theory that 
$$
\sqrt{N} (\hat{\Delta}_{HT} - \Delta) \rightarrow_d N(0,\Sigma_{HT}),
$$
where $\Sigma_{HT}$ is computed in \cite{lunceford2004stratification}. 
To prove Theorem 1, we connect  $\hat{\Delta}_{HT}^{S}$ and $\hat{\Delta}_{HT}$ with an intermediate (infeasible) estimator $\hat{\Delta}_{S-HT}$:
\begin{equation*}
	\label{eqn:ACE_S-IPW}
	\hat{\Delta}_{S-HT} = \sum\limits_{k=1}^K \left\{ \dfrac{1}{N{p_k}}  \sum\limits_{i=1}^N Z_i Y_i I(\hat{\pi}_i\in \hat{C}_k)    -
	\dfrac{1}{N(1-{p_k})}   \sum\limits_{i=1}^N (1-Z_i)Y_i I(\hat{\pi}_i \in \hat{C}_k)    \right\},
\end{equation*}
where $p_k = pr(Z=1\mid \hat{\pi}_i \in \hat{C}_k).$

In the first step, 
Lemma \ref{thm:ATE_S-IPW_consistent} shows that the difference between $\hat{\Delta}_{S-HT}$ and $\hat{\Delta}_{HT}$ tends to zero. We defer the proof of  Lemma \ref{thm:ATE_S-IPW_consistent}   to the end of this section. In the second step, we show that the difference between $\hat{\Delta}_{HT}^{S}$ and $\hat{\Delta}_{S-HT}$ tends to zero. 
%
%We first consider consistency of $\hat{\Delta}_{S-HT}$. Intuitively, as the number of subclasses goes to infinity and width of each interval $\hat{C}_k$ goes to zero, the subclass weights $p_{\hat{k}_i}$ converge to individual weights $\pi_i$. This idea is formalized in Lemma \ref{thm:ATE_S-IPW_consistent}.

\begin{lemma}
	\label{thm:ATE_S-IPW_consistent}
	Under  Assumption 1, the regularity conditions in the Supplementary Material,  and condition (3.4),
	\begin{itemize}
		\item [(i)] 	$ \hat{\Delta}_{S-HT}$ is consistent  for estimating $\Delta$:
		$  \hat{\Delta}_{S-HT} -  \Delta= o_p (1);$
		\item [(ii)]  If we assume additionally that (3.5) holds,
		then $ \hat{\Delta}_{S-HT}$ is $\sqrt{N}$-consistent  for estimating $\Delta$:
		$   \sqrt{N}      (   \hat{\Delta}_{S-HT} -  \Delta) = O_p (1).$
	\end{itemize} 
	
\end{lemma}

%We now introduce constraints on the rate of $K(N)$. As is obvious from equation (\ref{eqn:ATE_VS}), $\hat{\Delta}_S$ is only well-defined if each subclass contains observations from both treatment groups.  Larger number of subclasses results in fewer units in each subclass, and may lead to an ill-defined stratification estimator. Lemma \ref{lemma: ATE_VS_well} gives the condition for an asymptotically well-defined estimator:

We now turn to the second step, in which we show that under (3.5),
\begin{equation}
\label{eqn:ATE_VS-S-IPW}
\sqrt{N}   (  \hat{\Delta}_{HT}^{S} - \hat{\Delta}_{S-HT}  )  = O_p(1).
\end{equation}
%i.e.
%\begin{equation}
%	\label{eqn:diff_vs_vsipw_raw}
%	\sqrt{N}  \sum\limits_{k=1}^K \dfrac{1}{N} \left\{  \left( \dfrac{n_k}{n_{1k}} - \dfrac{1}{p_k}   \right)   \sum\limits_{i=1}^N Z_i Y_i I(\hat{\pi}_i \in \hat{C}_k)  \right\}= O_p(1),
%\end{equation}
By symmetry, we only show \eqref{eqn:ATE_VS-S-IPW} for the active treatment group, i.e.
\begin{equation}
\label{eqn:diff_vs_vsipw}
\sqrt{N} \sum\limits_{k=1}^K  \dfrac{n_k}{N} \left\{  \left( 1 - \dfrac{n_{1k}}{n_k p_k}   \right)  \dfrac{1}{n_{1k}} \sum\limits_{i=1}^N Z_i Y_i I(\hat{\pi}_i \in \hat{C}_k)  \right\}= O_p(1),
\end{equation}
where $K$ is used as a shorthand for $K(N)$. 

Without loss of generality,  we assume $n_1=\cdots=n_K=N/K \triangleq n$. 
%It can be shown that when $N$ is not a multiply of $K$, the remainder term tends to zero.
As $\hat{\Delta}_{HT}^{S}$ is well-defined, $n_{1k} \notin \{0,n\}, k=1, \ldots, K$. Thus $n_{1k}\sim tBin(n,p_k)$, where $tBin$ denotes truncated binomial distribution with range $[1,n-1]$.

{We will use Markov's inequality to show
(\ref{eqn:diff_vs_vsipw}).}  Denote
\begin{flalign*}
h_k &=  \dfrac{1-p_k^{n-1}}{1-p_k^n-(1-p_k)^n} - \dfrac{n_{1k}}{np_k}= c_{Nk} - \dfrac{n_{1k}}{np_k}\quad  \text{  (note }E[h_k]=0);\\
m_{1k} &= \dfrac{1}{n_{1k}} \sum\limits_{i=1}^N Z_i Y_i I(\hat{\pi}_i \in \hat{C}_k); \\
S_{NK} &=      \dfrac{1}{K} \sum\limits_{k=1}^{K} h_k m_{1k}.
\end{flalign*}
Then the left hand side of \eqref{eqn:diff_vs_vsipw} can be written as
\begin{equation}
\begin{aligned}
\label{eqn:diff_decomp}
& \sqrt{N} \sum\limits_{k=1}^K  \dfrac{n}{N} \left\{  \left( 1 - c_{Nk}   \right)  \dfrac{1}{n_{1k}} \sum\limits_{i=1}^N Z_i Y_i I(\hat{\pi}_i \in \hat{C}_k)  \right\}\\
&+\sqrt{N} \sum\limits_{k=1}^K  \dfrac{n}{N} \left\{  \left(c_{Nk}-\dfrac{n_{1k}}{n p_k}   \right)  \dfrac{1}{n_{1k}} \sum\limits_{i=1}^N Z_i Y_i I(\hat{\pi}_i \in \hat{C}_k)  \right\}\\
=& \dfrac{\sqrt{N}}{K}\sum\limits_{k=1}^K (1-c_{Nk}) m_{1k}+\sqrt{N}S_{NK}.
\end{aligned}
\end{equation}
We shall show that
\begin{equation}
\label{eqn:target1}
  \dfrac{\sqrt{N}}{K}\sum\limits_{k=1}^K (1-c_{Nk}) m_{1k}=o_p(1), 
\end{equation}
and 
\begin{equation}
\label{eqn:target2}
    \sqrt{N}S_{NK}=O_p(1).
\end{equation}

Let $\pi_{thres} = \min\{\pi_{min}/2, (1-\pi_{max})/2\}$. Note that equation \eqref{eqn:15} (see the proof of Lemma \ref{thm:ATE_S-IPW_consistent}) implies $c_{Nk}-1 =\{p_k^{n}-p_k^{n-1} + (1-p_k)^n\}/\{1-p_k^n - (1-p_k)^n\} = O((1-\pi_{thres})^n)$.
Moreover, by symmetry, we have
\begin{flalign*}
E(m_{1k}^2\mid n_{1k}) 
% 	&= 
% 	\dfrac{1}{n_{1k}^2}  E[(\sum_i Z_i Y_i I(\hat{\pi}_i \in \hat{C}_k))^2|n_{1k}]\\
% 	&=\dfrac{1}{n_{1k}^2} \left[   N E[Z_1 Y_1^2 I(\hat{\pi}_1 \in \hat{C}_k)|n_{1k}] + N(N-1) E[Z_1 Y_1^2 I(\hat{\pi}_1 \in \hat{C}_k) Z_2 Y_2^2 I(\hat{\pi}_2 \in \hat{C}_k)|n_{1k}]  \right]\\
% 	%  &=\dfrac{1}{n_{1k}^2} \left[   n_{1k} E[Y_1^2 |Z_1I(\hat{\pi}_1 \in \hat{C}_k)=1,n_{1k}] + n_{1k}(n_{1k}-1) E[Y_1 Y_2|Z_1  I(\hat{\pi}_1 \in \hat{C}_k) Z_2 I(\hat{\pi}_2 \in \hat{C}_k)=1,n_{1k}]  \right]\\
% 	&= \dfrac{1}{n_{1k}^2} \left[   n_{1k} E[Y_1^2 |Z_1I(\hat{\pi}_1 \in \hat{C}_k)=1] + n_{1k}(n_{1k}-1) E[Y_1 Y_2|Z_1  I(\hat{\pi}_1 \in \hat{C}_k) Z_2 I(\hat{\pi}_2 \in \hat{C}_k)=1]  \right]\\
% 	&\triangleq  \dfrac{1}{n_{1k}^2} \left[   n_{1k} \mu_{1k}^{11} + n_{1k}(n_{1k}-1) \mu_{1k}^{12}  \right]\\
&= \mu_{1k}^{12} + a_{2} \dfrac{1}{n_{1k}},
\end{flalign*}
where $a_{2}=\mu_{1k}^{11}-\mu_{1k}^{12}$ with   $\mu_{1k}^{11} = E\{Y_1^2 \mid Z_1I(\hat{\pi}_1 \in \hat{C}_k)=1\}$ and  $ \mu_{1k}^{12} = E\{Y_1 Y_2\mid Z_1  I(\hat{\pi}_1 \in \hat{C}_k) Z_2 I(\hat{\pi}_2 \in \hat{C}_k)=1\}$.
Furthermore, 
\[
E(m_{1k}^2)=E\left(\mu_{1k}^{12} + a_{2} \dfrac{1}{n_{1k}}\right)
=\mu_{1k}^{12} + a_{2}\left\{ \dfrac{np_k}{(np_k+r_k)^2}+O\left(\dfrac{1}{n^2}\right) \right\},
\]
where $r_k  = 1-p_k$, and we use the fact that
$E(1/n_{1k}) = np_k/\{(np_k + r_k)^2\} + O(1/n^2)$  \citep{znidaric2005asymptotic}. According to Assumption S5, $\mu_{1k}^{12}=O(1)$ and $a_2=O(1)$. Thus, $E(m_{1k}^2)=O(1)$. 
Based on the Markov's inequality, we have $m_{1k}=O_p(1)$. If Assumptions (3.5) and (3.6) are also satisfied,
then $\sqrt{N}\max_{1\leq k\leq K}|(1-c_{Nk})m_{1k}|=\sqrt{N}O((1-\pi_{thres})^n)O_p(1)=o_p(1)$. Since  $\sqrt{N}/{K}\sum_{k=1}^K (1-c_{Nk}) m_{1k}\leq \sqrt{N}\max_{1\leq k\leq K}|(1-c_{Nk})m_{1k}|$, then  \eqref{eqn:target1} holds. 

Note that
\begin{equation}
E(NS_{NK}^2)=E \dfrac{N}{K^2} \left(\sum\limits_{k=1}^{K} h_k m_{1k}  \right)^2 = \dfrac{N}{K^2}\left\{\sum_{k=1}^{K}E(h_k^2m_{1k}^2)+\sum_{k\neq j}E(h_km_{1k}h_jm_{1j}) \right\}.
\label{NS2}
\end{equation}
Under Assumptions S5, (3.5) and (3.6),
\begin{equation}
\begin{aligned}
\max\limits_{1\leq k\leq K}E(h_k^2m_{1k}^2)=&\max\limits_{1\leq k\leq K}E\{E(h_k^2m_{1k}^2\mid n_{1k})\}\\
=&\max\limits_{1\leq k\leq K}E\{h_k^2E(m_{1k}^2\mid n_{1k})\}\\
=&\max\limits_{1\leq k\leq K}E\left\{h_k^2 \left(\mu_{1k}^{12} + a_{2} \dfrac{1}{n_{1k}}\right)\right\}\\
=&\max\limits_{1\leq k\leq K}\left\{\mu_{1k}^{12}E\left(c_{Nk}^2-\dfrac{2c_{Nk}n_{1k}}{np_k}+\dfrac{n_{1k}^2}{n^2p_k^2}   \right)+a_2 E\left(\dfrac{c_{Nk}^2}{n_{1k}}-\dfrac{2c_{Nk}}{np_k}+\dfrac{n_{1k}}{n^2p_k^2}   \right)\right\}\\
=&\max\limits_{1\leq k\leq K}\Bigg[\mu_{1k}^{12}\left\{c_{Nk}^2-2c_{Nk}^2+\dfrac{(n^2p_k^2+np_k-np_k^2-n^2p_k^n)c_{Nk}}{(1-p_k^{n-1})n^2p_k^2}\right\}\\
&\qquad \quad +a_2 \left[c_{Nk}^2\left\{ \dfrac{np_k}{(np_k+r_k)^2}+O\left(\dfrac{1}{n^2}\right) \right\}-\dfrac{2c_{Nk}}{np_k}+\dfrac{c_{Nk}}{np_k}   \right]\Bigg]\\
=&\max\limits_{1\leq k\leq K}\Bigg[\left\{\dfrac{\mu_{1k}^{12}(n^2p_k^2+np_k-np_k^2-n^2p_k^n)}{(1-p_k^{n-1})n^2p_k^2}-\dfrac{(\mu_{1k}^{11}-\mu_{1k}^{12})}{np_k}     \right\}c_{Nk}\\
&\qquad\quad +\left\{(\mu_{1k}^{11}-\mu_{1k}^{12})\dfrac{np_k}{(np_k+r_k)^2}-\mu_{1k}^{12}\right\}c_{Nk}^2+O\left(\dfrac{1}{n^2}\right)\Bigg]\\
=&O\left(\dfrac{1}{n}\right)+O\left(\dfrac{1}{n^2}\right)=O\left(\dfrac{1}{n}\right),
\end{aligned}
\label{part1}
\end{equation}
where the last equality holds because $p_k$ is bounded away from 0 and 1.
Moreover,
\begin{equation}
\begin{aligned}
&\max\limits_{k,j\in\{1,\ldots,K\},k\neq j}E(h_km_{1k}h_jm_{1j})\\
=&\max\limits_{k,j\in\{1,\ldots,K\},k\neq j}E\{E(h_km_{1k}h_jm_{1j}\mid n_{1k},n_{1j})\}\\
=&\max\limits_{k,j\in\{1,\ldots,K\},k\neq j}E\{h_kh_jE(m_{1k}m_{1j}\mid n_{1k},n_{1j})\}\\
=&\max\limits_{k,j\in\{1,\ldots,K\},k\neq j}E\left[h_kh_jE\left\{\dfrac{1}{n_{1k}} \sum\limits_{i=1}^N Z_i Y_i I(\hat{\pi}_i \in \hat{C}_k)\dfrac{1}{n_{1j}} \sum\limits_{i=1}^N Z_i Y_i I(\hat{\pi}_i \in \hat{C}_j)\big\mid n_{1k},n_{1j}\right\}\right]\\
=&\max\limits_{k,j\in\{1,\ldots,K\},k\neq j}E\left[h_kh_jE\left\{ Y_1  Y_{2}\big\mid Z_1  I(\hat{\pi}_1 \in \hat{C}_k)Z_{2}  I(\hat{\pi}_{2} \in \hat{C}_j)=1 \right\}\right] \text{  (by symmetry)} \\
=&\max\limits_{k,j\in\{1,\ldots,K\},k\neq j}E\left\{\left(c_{Nk} - \dfrac{n_{1k}}{np_k}\right)\left(c_{Nj} - \dfrac{n_{1j}}{np_j}\right)\right\}E\left\{ Y_1  Y_{2}\big\mid Z_1  I(\hat{\pi}_1 \in \hat{C}_k)Z_{2}  I(\hat{\pi}_{2} \in \hat{C}_j)=1 \right\}\\
=&\max\limits_{k,j\in\{1,\ldots,K\},k\neq j}\frac{1}{n^2p_kp_j}\left\{E(n_{1k}n_{1j})-E(n_{1k})E(n_{1j})\right\}E\left\{ Y_1  Y_{2}\big\mid Z_1  I(\hat{\pi}_1 \in \hat{C}_k)Z_{2}  I(\hat{\pi}_{2} \in \hat{C}_j)=1 \right\}\\
=&\max\limits_{k,j\in\{1,\ldots,K\},k\neq j}\frac{1}{n^2p_kp_j}\text{cov}(n_{1k},n_{1j})E\left\{ Y_1  Y_{2}\big\mid Z_1  I(\hat{\pi}_1 \in \hat{C}_k)Z_{2}  I(\hat{\pi}_{2} \in \hat{C}_j)=1 \right\}\\
=&\max\limits_{k,j\in\{1,\ldots,K\},k\neq j}\Bigg[\frac{1}{n^2p_kp_j}\Big[E\{\text{cov}(n_{1k},n_{1j}\mid I(\hat{\pi}_i\in \hat{C}_\ell),i=1,\ldots,N,\ell=1,\ldots,K)\}\\
&\!+\!\text{cov}\left\{E\left(n_{1k}\!\mid\! I(\hat{\pi}_i\in \hat{C}_\ell),i=1,\ldots,N,\ell=1,\ldots,K\right),  E(n_{1j}\!\mid\! I(\hat{\pi}_i\in \hat{C}_\ell),i=1,\ldots,N,\ell=1,\ldots,K) \right\}\Big]\\
&\qquad \qquad\qquad \times E\left\{ Y_1  Y_{2}\big\mid Z_1  I(\hat{\pi}_1 \in \hat{C}_k)Z_{2}  I(\hat{\pi}_{2} \in \hat{C}_j)=1 \right\}\Bigg]\\
=&\max\limits_{k,j\in\{1,\ldots,K\},k\neq j}\Bigg[\frac{1}{n^2p_kp_j}E\{\text{cov}(n_{1k},n_{1j}\mid I(\hat{\pi}_i\in \hat{C}_\ell),i=1,\ldots,N,\ell=1,\ldots,K)\}\\
&\qquad\qquad \qquad\times E\left\{ Y_1  Y_{2}\big\mid Z_1  I(\hat{\pi}_1 \in \hat{C}_k)Z_{2}  I(\hat{\pi}_{2} \in \hat{C}_j)=1 \right\}\Bigg]\\
\leq &\max\limits_{k,j\in\{1,\ldots,K\},k\neq j}\left[\frac{1}{n^2p_kp_j}\Big|E\left\{ Y_1  Y_{2}\big\mid Z_1  I(\hat{\pi}_1 \in \hat{C}_k)Z_{2}  I(\hat{\pi}_{2} \in \hat{C}_j)=1 \right\}\Big|\right]\\
&\times E\left\{\max\limits_{k,j\in\{1,\ldots,K\},k\neq j}\Big|\text{cov}(n_{1k},n_{1j}\mid I(\hat{\pi}_i\in \hat{C}_\ell),i=1,\ldots,N,\ell=1,\ldots,K)\Big|\right\}.\\
\end{aligned}
\label{part2}
\end{equation}
Under Assumption S5, we have $\big|E\left\{ Y_1  Y_{2}\big\mid Z_1  I(\hat{\pi}_1 \in \hat{C}_k)Z_{2}  I(\hat{\pi}_{2} \in \hat{C}_j)=1 \right\}\big|=O(1)$. Combined with Assumption S6, we then get that
\begin{equation}
\label{eqn:new}
\max_{k,j\in\{1,\ldots,K\},k\neq j}E(h_km_{1k}h_jm_{1j})=O(1/N).
\end{equation}

Equations \eqref{NS2}, \eqref{part1} and \eqref{eqn:new} together imply that $E(NS_{NK}^2)=O(1)$. Based on the Markov's inequality, we can get \eqref{eqn:target2}.

We have hence finished the proof.

\section{Proof of Lemma \ref{thm:ATE_S-IPW_consistent}}
\label{sec:proof_S-IPW_consistent}

For simplicity we only prove claim (ii). Proof of claim (i) can be obtained following similar arguments.
Due to the asymptotic normality of $\hat{\Delta}_{HT}$,  it suffices to show that
$$
\sqrt{N}      (   \hat{\Delta}_{S-HT} - \hat\Delta_{HT} ) = O_p(1).
$$

Let $\pi_1,\ldots,\pi_n$ be independent samples of $\pi(\bm X;\beta)$, and $F^{-1}(\cdot)$ be the quantile distribution of $\pi(\bm X;\beta)$.  For $t\in (0,1)$, the empirical quantile distribution is defined as
$$
\mathbb{F}_N^{-1}(t) = \inf\left\{x:\mathbb{F}_N (x) \geq t \right\},
$$
where $\mathbb{F}_N(x) = \sum_{i=1}^N 1_{(-\infty,x]}(\pi_i)   / N$ is the empirical distribution function. 
% When no confusion is possible, we use $K$ as a shorthand for $K(N)$.

Using standard empirical process theory \citep{shorack2009empirical}, we can show that
\begin{equation}
\label{eqn:conv-quantile}
\sqrt{N}\Vert \mathbb{F}_N^{-1} - F^{-1} \Vert_0^1 =\sqrt{N} \sup\limits_{0\leq t\leq 1} |\mathbb{F}_N^{-1}(t) -  F^{-1}(t)| = O_p(1).
\end{equation}
As $F^{-1}(t)$ is Lipschitz continuous,
\begin{equation}
\label{eqn:abso-cont}
K \max\limits_{1\leq k \leq K} \{q_k-q_{k-1}\} = O(1),
\end{equation}
where $q_k = F^{-1}\left(k/K \right)$. Assumption (3.5), results (\ref{eqn:conv-quantile}) and (\ref{eqn:abso-cont}) together imply
\begin{equation}
\label{eqn:3}
\sqrt{N} \max\limits_{1\leq k \leq K} |\tilde{q}_k- q_k| =O_p(1) \text{\quad and \quad}
\sqrt{N} \max\limits_{1\leq k \leq K} \{\tilde{q}_k-\tilde{q}_{k-1}\} =O_p(1),
\end{equation}
where $\tilde{q}_k = \mathbb{F}_N^{-1}\left(k/K \right) \ (k=1,\ldots,K)$, the sample quantiles of the (true) propensity scores. Now let $\hat{q}_k$ be the sample quantiles of the estimated propensity scores, Assumptions (3.5) and \ref{assump:correct_PS},  and result (\ref{eqn:3}) imply
\begin{equation}
\label{eqn:qk-close}
\sqrt{N}\max\limits_{1\leq k \leq K} |\hat{q}_k-q_k| =O_p(1) \text{\quad and \quad} \sqrt{N}\max\limits_{1\leq k \leq K} \{\hat{q}_k-\hat{q}_{k-1}\} =O_p(1).
\end{equation}
Denote $\hat{\pi}_{min}=\min\{\hat{\pi}_i, i=1,\ldots,N\}$, $\hat{\pi}_{max}=\max\{\hat{\pi}_i, i=1,\ldots,N\}$ and $\pi_{thres} = \min\{\pi_{min}/2, (1-\pi_{max})/2\}$, Assumptions \ref{assump:positivity} and \ref{assump:correct_PS} imply that for large enough $N$,
\begin{equation}
\label{eqn:13}
\pi_{thres} < \min(\hat{\pi}_{min},  1-\hat{\pi}_{max}).
\end{equation}
Moreover, if we let $\delta = \max_{1\leq i\leq N} |\hat{\pi}_{i}-\pi_i|$, then
\begin{flalign*}
p_{\hat{k}_i}  &=  P(Z=1\mid \hat{\pi}_i \in [\hat{q}_{\hat{k}_i-1}, \hat{q}_{\hat{k}_i}]) \\
&=E_{\pi_i\mid \hat{\pi}_i \in [\hat{q}_{\hat{k}_i-1}, \hat{q}_{\hat{k}_i}]} E\left\{ P(Z=1\mid \pi_i,  \hat{\pi}_i \in [\hat{q}_{\hat{k}_i-1}, \hat{q}_{\hat{k}_i}]) \mid \hat{\pi}_i \in [\hat{q}_{\hat{k}_i-1}, \hat{q}_{\hat{k}_i}] \right\} \\
&=E_{\pi_i\mid \hat{\pi}_i \in [\hat{q}_{\hat{k}_i-1}, \hat{q}_{\hat{k}_i}]} E\left\{ \pi_i \mid \hat{\pi}_i \in [\hat{q}_{\hat{k}_i-1}, \hat{q}_{\hat{k}_i}] \right\} \\
&\in [\hat{q}_{\hat{k}_i-1} - \delta, \hat{q}_{\hat{k}_i} + \delta]. 
\end{flalign*}
On the other hand,  $\hat{\pi}_i \in [\hat{q}_{\hat{k}_i-1}, \hat{q}_{\hat{k}_i}]  \subset [\hat{q}_{\hat{k}_i-1} - \delta, \hat{q}_{\hat{k}_i} + \delta]$, hence 
\begin{equation}
\label{eqn:decom}
|p_{\hat{k}_i} - \hat{\pi}_i|  \leq 
\max\limits_{1\leq k \leq K} \{\hat{q}_{k}-\hat{q}_{k-1}\}+ 2\max\limits_{1\leq i\leq N} |\hat{\pi}_{i}-\pi_i|.
\end{equation}
Combining \eqref{eqn:decom} with \eqref{eqn:qk-close}, \eqref{eqn:13},   and Assumption \ref{assump:correct_PS}, we have for large enough $N$, 
\begin{equation}\label{eqn:15}
\pi_{thres} \leq p_k \leq 1-\pi_{thres}.
\end{equation}

On the other hand, 
for large enough $N$, we have
\begin{flalign*}
&\sqrt{N}| \hat{\Delta}_{S-HT}-\hat{\Delta}_{HT} |\\ &\leq            \sqrt{N} \dfrac{1}{N} \sum\limits_{i=1}^N |Y_i| I(Z_i=1)
\left| \dfrac{1}{p_{\hat{k}_i}} - \dfrac{1}{\hat{\pi}_i}\right|
+            \sqrt{N} \dfrac{1}{N} \sum\limits_{i=1}^N |Y_i| I(Z_i=0) \left|\dfrac{1}{1-p_{\hat{k}_i}} - \dfrac{1}{1-\hat{\pi}_i}\right| \\
& \leq            \sqrt{N} \dfrac{1}{N} \sum\limits_{i=1}^N |Y_i|  \dfrac{ |p_{\hat{k}_i} - \hat{\pi}_i| }{\pi_{thres}^2} \\
&\leq           \sqrt{N}  \dfrac{1}{N} \sum\limits_{i=1}^N |Y_i|  \dfrac{ \max\limits_{1\leq k \leq K} \{\hat{q}_{k}-\hat{q}_{k-1}\}+ 2\max\limits_{1\leq i\leq N} |\hat{\pi}_{i}-\pi_i| }{\pi_{thres}^2}   \\
& =O_p(1).
\end{flalign*}
%where the second to the last line follows from the fact that both $p_{\hat{k}_i}$ and $\hat{\pi}_i$ fall in the interval $\hat{C}_{\hat{k}_i}$.
Hence we complete the proof of Lemma \ref{thm:ATE_S-IPW_consistent}.

\section{Proof of Theorem 2}
\label{sec:proof_VS_well}

When $N$ is large enough, by uniform convergence of $\hat{\pi}_i \ (i=1,\ldots,N)$ and uniform convergence of sample quantiles $\hat{q}_k \ (k=1,\ldots,K)$ (see Section \ref{sec:proof_S-IPW_consistent} for the detailed proof), we have for large enough $N$,
\begin{flalign*}
{\pi_{thres}} \leq pr(Z=1\mid \hat{\pi}(\bm X) \in \hat{C}_k)  \leq 1-{\pi_{thres}}.
\end{flalign*}
Then
\begin{flalign*}
&\quad \ \  pr(\text{exists } z,k, \text{ such that } n_{zk}=0)\\ &\leq \sum\limits_{z=0}^1 \sum\limits_{k=1}^K  pr(n_{zk}=0) \\
&\leq  \sum\limits_{k=1}^K  \left( pr(Z=1\mid \hat{\pi}_i \in \hat{C}_k)^{n_k} +  pr(Z=0\mid \hat{\pi}_i \in \hat{C}_k)^{n_k} \right) \\
&\leq \sum\limits_{k=1}^K   2(1-{\pi_{thres}})^{N/K - 1}\\
&= \exp\left\{  \log(2K) +  \left(\dfrac{N}{K} - 1\right) \log(1-\pi_{thres})     \right\} \\
& \rightarrow  0.
\end{flalign*}
This completes the proof of Theorem 2.

\section{Proof that $K_{max}$ satisfies the rate condition in Theorem 1}

Suppose that $K(N)$ grows at a polynomial rate slower than 1, i.e. $K(N)= N^\alpha, \alpha<1$, then 
$$
K(N)\log(K(N)) / N = \alpha \log N / N^{1-\alpha} \rightarrow 0
$$
as $N\rightarrow \infty$. Hence by Theorem 2, $\hat{\Delta}_{HT}^S$ is asymptotically well-defined.  {Now pick any $\alpha_0 \in (0.5,1),$ say $\alpha_0=0.9$. From the definition of $K_{\text{max}}$ we have that $N^{0.9}/K_{\text{max}} = O(1)$.} 
% \yz{YZ: why it is satisfied? Based on the definition of $K_{max}$, $K_{max}$ is the largest value of $K$ so that condition of $K(N)\log(K(N)) / N\rightarrow 0$ can be satisfied. Then $K_{max}$ can be taken as $N^{\alpha_1}$, where $\alpha_1$ can be very close to 1, so $\alpha_1>\alpha_0=0.9$ and $N^{0.9}/K_{\text{max}} = o(1)$, not $O(1)$. } 
Consequently, 
$$
K_{\text{max}} / \sqrt{N} \rightarrow \infty
$$
as $N\rightarrow \infty$. This is exactly equation (3.5) in Theorem 1.

\section{Proof  of Theorem 3}

Let $$
\hat{\Delta}_{DR}^{S} =  \dfrac{1}{N} \sum\limits_{i=1}^N \dfrac{Z_i Y_i - (Z_i-\hat{p}_{\hat{k}_i}) \hat{b}_1(\bm X_i)}{\hat{p}_{\hat{k}_i}} - \dfrac{1}{N} \sum\limits_{i=1}^N \dfrac{(1-Z_i) Y_i + (Z_i-\hat{p}_{\hat{k}_i}) \hat{b}_0(\bm X_i)}{1-\hat{p}_{\hat{k}_i}} \equiv \hat{\Delta}_{DR,1}^{S} - \hat{\Delta}_{DR,0}^{S}.
$$
Denote $B_1^*$ and $\Pi^*$ as the probability limit of $\hat{B}_1$ and $\widehat{\Pi}$, respectively.
We now show that the bias of $\hat{\Delta}_{DR,1}^S$ for estimating $\Delta_1 \equiv E\{Y(1)\}$ admits a product structure. Results for $\hat{\Delta}_{DR,0}^S$ can be shown similarly and are omitted.
\begin{flalign*}
\hat{\Delta}_{DR,1}^S - \Delta_1 =& \mathbb{P}_N \left(\dfrac{Z(Y-\hat{B}_1)}{\widehat{\Pi}} + \hat{B}_1 \right)  - \Delta_1 \\
=& (\mathbb{P}_N-\mathbb{P}_0) \left(\dfrac{Z(Y-\hat{B}_1)}{\widehat{\Pi}} + \hat{B}_1 \right)+ 
\mathbb{P}_0 \left(\dfrac{Z(Y-\hat{B}_1)}{\widehat{\Pi}} + \hat{B}_1 \right)  - \Delta_1\\
=& (\mathbb{P}_N-\mathbb{P}_0) \left(\dfrac{Z(Y-\hat{B}_1)}{\widehat{\Pi}} + \hat{B}_1 - \dfrac{Z(Y-{B}^*_1)}{{\Pi}^*} - {B}^*_1 \right) + \\
&  (\mathbb{P}_N-\mathbb{P}_0) \left(\dfrac{Z(Y-{B}^*_1)}{{\Pi}^*} + {B}_1^* \right)+ 
\mathbb{P}_0 \left(\dfrac{Z(Y-\hat{B}_1)}{\widehat{\Pi}} + \hat{B}_1 \right)  - \Delta_1 \\
=& o_p\left(N^{-1/2}\right) +   (\mathbb{P}_N-\mathbb{P}_0) \left(\dfrac{Z(Y-{B}^*_1)}{{\Pi}^*} + {B}^*_1 \right) +	\mathbb{P}_0 \left(\dfrac{Z(Y-\hat{B}_1)}{\widehat{\Pi}} + \hat{B}_1 \right)   - \Delta_1\\
\equiv & o_p\left(N^{-1/2}\right)  +   (\mathbb{P}_N-\mathbb{P}_0) D_{dr}({B}^*_1,\Pi^*) + \mathbb{P}_0 \left(\dfrac{Z(Y-\hat{B}_1)}{\widehat{\Pi}} + \hat{B}_1 \right)   - \mathbb{P}_0 \left(B_1 \right) \\
=& o_p\left(N^{-1/2}\right) +   (\mathbb{P}_N-\mathbb{P}_0) D_{dr}({B}^*_1,\Pi^*) +  \mathbb{P}_0 \left(\dfrac{(\Pihat-\Pi) (\hat{B}_1-B_1)}{\widehat{\Pi}}\right), \numberthis\label{eqn:neg}
\end{flalign*}
where $\mathbb{P}_N$  denotes the empirical mean: $\mathbb{P}_N(O) =  \sum_{i=1}^{N} O_i/N$, and $\mathbb{P}_0$ denotes  expectation conditional on estimated functionals $\hat{\pi}(\cdot)$ and $\hat{b}(\cdot)$. 
Due to the Cauchy-Schwartz inequality and equation \eqref{eqn:13}, the last term in \eqref{eqn:neg} satisfies
\begin{equation}
\label{eqn:cs}
\mP_0 \left\{ \dfrac{ (\widehat{\Pi} - \Pi)(\hat{B}_1 - B_1) }{\widehat{\Pi}} \right\} \leq  \dfrac{1}{\pi_{thres}}
\mP_0\Vert\widehat{\Pi} - \Pi \Vert \mP_0\Vert \hat{B}_1 - B_1 \Vert.
\end{equation}
The fourth equality in \eqref{eqn:neg} holds if $D_{dr}(\hat{B}_1, \widehat{\Pi})$ falls in a Donsker class with probability tending to 1, and $\mP_0[ \{D_{dr}(\hat{B}_1, \widehat{\Pi})  - D_{dr}(B_1^{*}, \Pi^{*})\}^2] \rightarrow_p 0$.

We now discuss the convergence property of $\Vert\widehat{\Pi} - \Pi \Vert$. Note that by the triangle inequality,
\begin{equation}
\begin{aligned}
\mP_N\Vert\widehat{\Pi} - \Pi \Vert &\leq \left\{ \dfrac{1}{N} \sum\limits_{i=1}^N (p_{\hat{k}_i} - \hat{p}_{\hat{k}_i})^2  \right\}^{1/2} + \left\{\dfrac{1}{N}\sum\limits_{i=1}^N (p_{\hat{k}_i} - \hat{\pi}_i )^2 \right\}^{1/2} + \left\{\dfrac{1}{N}\sum\limits_{i=1}^N (\hat{\pi}_i - \pi_i)^2  \right\}^{1/2} \\
&\leq \left\{\dfrac{1}{N}\sum\limits_{k=1}^K n \left(\dfrac{n_{1k}}{n} -p_k\right)^2  \right\}^{1/2} + 
O_p\left(\dfrac{1}{K}\right) + O_p\left(\dfrac{1}{\sqrt{N}}\right) + O_p\left(\dfrac{1}{\sqrt{N}}\right)  \\
&= O_p\left(\dfrac{\sqrt{K}}{{\sqrt{N}}}\right) + O_p\left(\dfrac{1}{K}\right) + O_p\left(\dfrac{1}{\sqrt{N}}\right) \\
&= o_p(1),
\label{eqn:converge}
\end{aligned}
\end{equation}
where the second inequality is due to \eqref{eqn:decom}. Combining \eqref{eqn:converge} with \eqref{eqn:neg} and \eqref{eqn:cs}, we finish the proof. 

% \doubt{We also note that when $K = C \times {N}^{1/3}$ where $C$ is a constant,} \yz{YZ: $K$ can not take this value, because $K$ should satisfy the condition of $K/\sqrt{N}\rightarrow \infty$ as $N\rightarrow \infty$.} we have 
% $$
% \Vert\widehat{\Pi} - \Pi \Vert \leq \yz{O_p\left(\dfrac{\sqrt{K}}{{\sqrt{N}}}\right) }+ O_p\left(\dfrac{1}{K}\right) + O_p\left(\dfrac{1}{\sqrt{N}}\right) = O_p \left(\dfrac{1}{N^{1/3}}\right).$$
% In this case, condition (i) in Theorem 3 can be relaxed so that $	\Vert \hat{b}_z(\bm X_i) - b_z(\bm X_i)\Vert = o_p(1/{N^{1/6}}), z=0,1.$  As this convergence rate is much slower than the parametric convergence rate $O_p(1/\sqrt{N})$, this choice of $K$ allows for  use of off-the-shelf nonparametric regression and flexible machine learning tools for estimating
% the outcome regressions $b_z (X),z = 0,1.$ It remain a practical problem, however, to choose the constant $C$ in practice. We leave this as a future topic for investigation.

\section{Additional simulation results}
\label{appendix:addsimu}
\begin{figure}[ht]
	\centering
	\includegraphics[width=\linewidth]{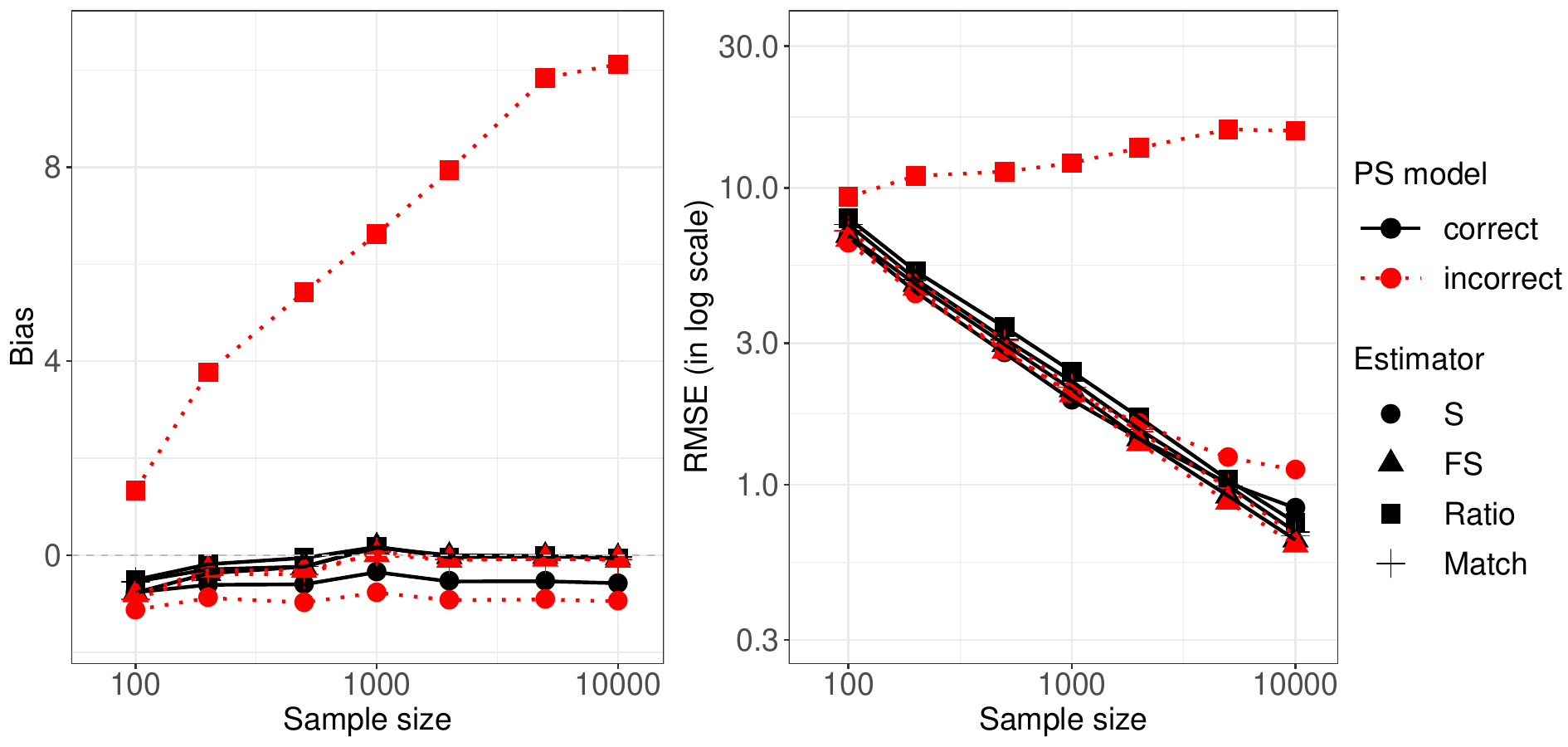}	
	\caption{Bias and root mean squared error (RMSE) of the classical subclassification estimator (S), the full subclassification estimator (FS), the Ratio estimator (Ratio) and the matching estimator (Match) under  \cite{chan2016globally}'s setting. 
	}
	\label{fig:simu_fs}
\end{figure}

\begin{figure}[ht]
	\centering
	\includegraphics[width=\linewidth]{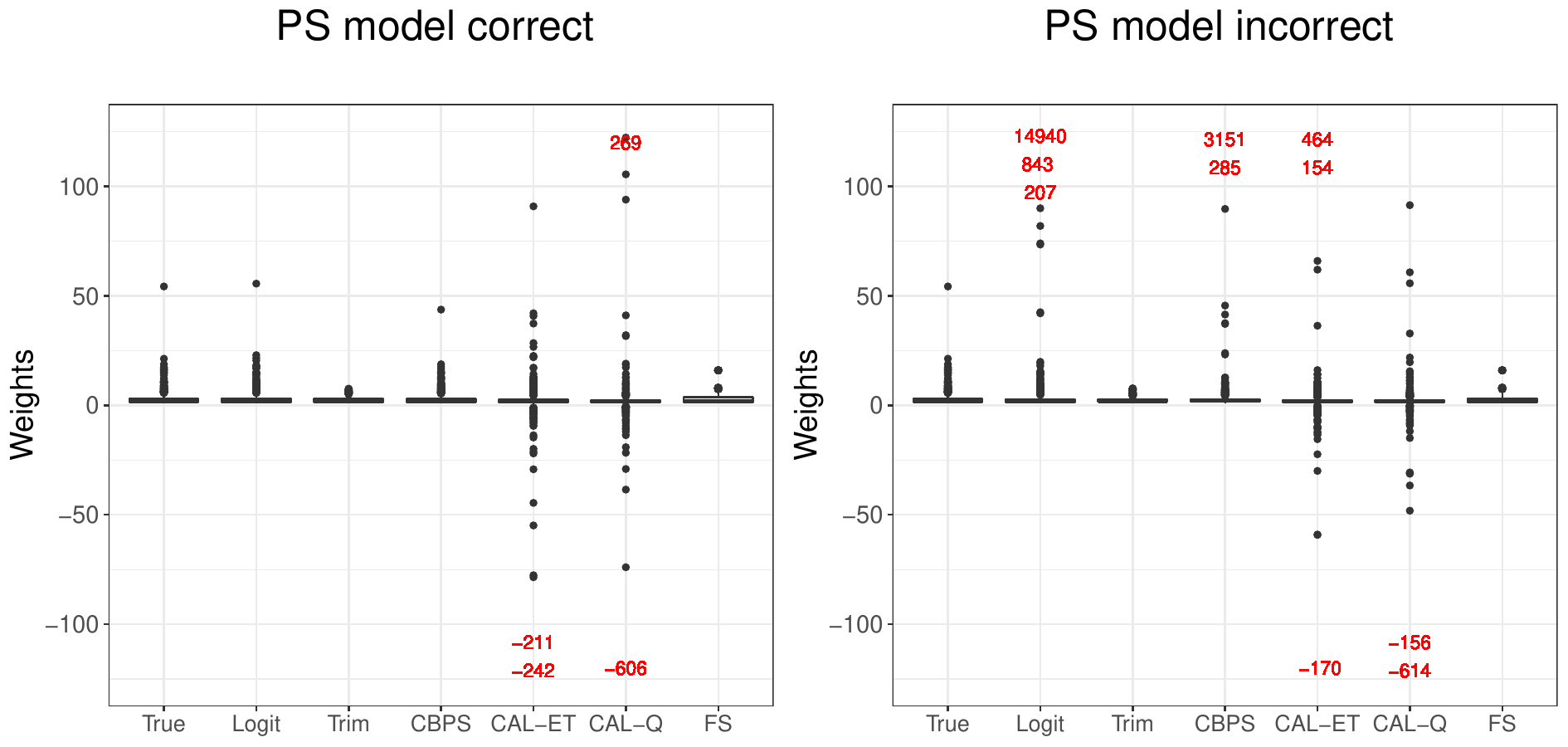}	
	\caption{Distributions of weight estimates with various weighting scheme with a random simulated data set of sample size 1000. Weights outside of the plot range are annotated on the boarders.}
	%		
	%		ATE estimates obtained with $\hat{\Delta}_{DR}^{CAL-ET}$, $\hat{\Delta}_{DR}^{CAL-Q}$ and $\hat{\Delta}_{DR}^{FS}$. The outcome regression model is misspecified for both plots; the propensity score model is correctly specified for the left panel, and misspecified for the right panel.  The horizontal line at zero corresponds to the true value of the ATE.  
	%	}
	\label{fig:simu_ks_boxplots}
\end{figure}

In this section, we present  additional results for the simulation studies. Figure \ref{fig:simu_fs} shows the figure corresponding to Table 1 in the main paper. Figure 	\ref{fig:simu_ks_boxplots} visualizes the weight stability of various weighting schemes; see also Figure 1 in the main paper.

\section{Descriptive statistics for the NHANES data}
\label{appendix:table1_nhanes}

Following \cite{chan2016globally}, we control for the following potential confounders in our analysis: child age, child gender, child race (black, Hispanic versus others), coming from a family above
200\% of the federal poverty level, participation in Special Supplemental Nutrition (SSN) Program for Women
Infants and Children, participation in the Food Stamp Program, childhood food security as measured by an indicator of two or more affirmative responses to eight child-specific questions in the NHANES
Food Security Questionnaire Module, any health insurance coverage, and the age and gender of the survey respondent (usually an adult in the family).

In Table \ref{tab:descriptives} we provide summary statistics for potential confounders and outcome measure by participation status in the school meal programs.

% latex table generated in R 3.1.2 by xtable 1.7-4 package
% Fri Dec 25 13:58:25 2015
\begin{table}[t!]
	\centering
	\caption{Baseline characteristics and outcome measure by participation status in the school meal programs}
	\label{tab:descriptives}
	\bigskip
	\begin{tabular}{rcc}
		\toprule
		& Participated  & Not participated \\ 
		&  (N=1284)   & (N=1046) \\ 
		\midrule
		Child Age, mean (SD) & 10.1 (3.5) & 9.9 (4.4) \\ 
		Child Male, N (\%) & 657 (51.2\%) & 549 (52.5\%) \\ 
		Black, N (\%) & 396 (30.8\%) & 208 (19.9\%) \\ 
		Hispanic, N (\%) & 421 (32.8\%) & 186 (17.8\%) \\ 
		Above 200\% of poverty level, N (\%) & 317 (24.7\%) & 692 (66.2\%) \\ 
		Participation in SSN program, N (\%) & 328 (25.5\%) & 115 (11.0\%) \\ 
		Participation in food stamp program, N (\%) & 566 (44.1\%) & 122 (11.7\%) \\ 
		Childhood food security, N (\%) & 418 (32.6\%) & 155 (14.8\%) \\ 
		Insurance Coverage, N (\%) & 1076 (83.8\%) & 927 (88.6\%) \\ 
		Respondent Age, mean (SD) & 38.6 (10.4) & 40.3 (9.7) \\ 
		Respondent Male, N (\%) & 506 (39.4\%) & 526 (50.3\%) \\ 
		BMI, mean (SD) & 20.4 (5.5) & 19.8 (5.4) \\ 
		\bottomrule
	\end{tabular}
\end{table}

\end{document}